\documentclass[a4paper,onecolumn,11pt,preprint]{quantumarticle}
\pdfoutput=1
\usepackage{mathtools}
\usepackage[utf8x]{inputenc}     
\usepackage{enumerate}
\usepackage{epsfig}     
\usepackage{float} 
\usepackage{subfig}
\usepackage{wasysym} 
\usepackage{mathrsfs} 
 \usepackage[colorlinks=true,     linkcolor=blue, citecolor=blue, filecolor=blue, urlcolor=blue]{hyperref}
\usepackage{amsfonts} 
\usepackage{amsbsy} 
\usepackage{amscd} 
\usepackage{pstricks} 
\usepackage{multirow} 
\usepackage{tikz}
\usepackage{color}
\usepackage{slashed}
\usepackage{array}
\usepackage{tablefootnote}

\usetikzlibrary{arrows,positioning,shapes.geometric} 
\usepackage[compat=1.1.0]{tikz-feynman}          
\usepackage[font=small,labelfont=bf]{caption}
\usepackage{slashed}
\usepackage{multirow}
\usepackage{tabularx}
\usepackage{soul}  
\usepackage{listings} 
\newif\iflocaldraftfigures
\localdraftfiguresfalse
\iflocaldraftfigures
\renewcommand{\includegraphics}[2][]{\fbox{\scriptsize\ttfamily\detokenize{#2}}}
\fi
\usepackage{color}

\usepackage{amsthm}

\newtheorem{definition}{Definition}

\newtheorem{observation}{Observation}
\newtheorem{proposition}{Proposition}

\newcommand{\w}{\mathbf{w}}
\newcommand{\W}{\mathbb{R}^{d_w}}

\newcommand{\X}{\mathcal{X}}

\newcommand{\M}{\mathcal{M}} 
\newcommand{\x}{\mathbf{x}}
\newcommand{\mf}[1]{\mathfrak{#1}}
\newcommand{\mc}[1]{\mathcal{#1}} 
\newcommand{\mbf}[1]{\mathbf{#1}}

\usepackage{makecell}
\usepackage{pifont}
\usepackage{amsthm}

\newcommand{\ket}[1]{|#1\rangle}
\newcommand{\bra}[1]{\langle#1|}
\newcommand{\braket}[2]{\langle#1|#2\rangle}
\newcommand{\quantexp}[2]{\langle#2|#1|#2\rangle}

\newcommand{\DJac}[1]{\text{rank}(\mbf{J}_\X(#1))}

\newcommand{\GL}[2]{\text{GL}(#1,#2)}
\newcommand{\din}{d_{\text{inp.}}}
\newcommand{\dX}{d_{\X}} 
\usepackage{amsthm}
\usepackage{booktabs}

\usepackage{mathtools}

\newcommand{\cmark}{\ding{51}}%
\newcommand{\xmark}{\ding{55}}%
\usetikzlibrary{quantikz}

\begin{document} 
\title{From Reachability to Learnability: Geometric Design Principles for Quantum Neural Networks} 

\author{Vishal~S.~Ngairangbam}
\email{vishal.s.ngairangbam@durham.ac.uk}
\affiliation{Institute for Particle Physics Phenomenology, Department of Physics \\
Durham University, Durham DH1 3LE, United Kingdom}

\author{Michael~Spannowsky}
\email{michael.spannowsky@durham.ac.uk}
\affiliation{Institute for Particle Physics Phenomenology, Department of Physics \\
Durham University, Durham DH1 3LE, United Kingdom}
\affiliation{Institute for Theoretical Physics, Karlsruhe Institute of Technology (KIT), \\ Wolfgang-Gaede-Str. 1, 76131 Karlsruhe, Germany}

	\maketitle

\begin{abstract}
Classical deep networks are effective because depth enables adaptive geometric deformation of data representations. In quantum neural networks (QNNs), however, depth or state reachability alone does not guarantee this feature-learning capability. We study this question in the pure-state setting by viewing encoded data as an embedded manifold in $\mathbb{C}P^{2^n-1}$ and analysing infinitesimal unitary actions through Lie-algebra directions. We introduce Classical-to-Lie-algebra (CLA) maps and the criterion of almost Complete Local Selectivity (aCLS), which combines directional completeness with data-dependent local selectivity. Within this framework, we show that data-indepen-dent trainable unitaries are complete but non-selective, i.e. learnable rigid reorientations, whereas pure data encodings are selective but non-tunable, i.e. fixed deformations. Hence, geometric flexibility requires a non-trivial joint dependence on data and trainable weights. We further show that accessing high-dimensional deformations of many-qubit state manifolds requires parametrised entangling directions; fixed entanglers such as CNOT alone do not provide adaptive geometric control. Numerical examples validate that aCLS-satisfying data re-uploading models outperform non-tunable schemes while requiring only a quarter of the gate operations. Thus, the resulting picture reframes QNN design from state reachability to controllable geometry of hidden quantum representations.
\end{abstract}

\tableofcontents
\flushbottom	

\section{Introduction}
Deep neural networks are powerful because each layer can reshape the geometry of data, not just re-label it. Quantum circuits can also be made deep, but depth alone does not guarantee this kind of adaptive feature learning. The central question of this work is therefore geometric:\footnote{With geometric deep learning's established paradigm in classical representation learning, parallel interest targets group-equivariant quantum models~\cite{Meyer:2022fjx,Das:2023ded,Tuysuz:2024gnk,Wiersema:2024qky}. We instead examine quantum circuits' intrinsic geometric flexibility: their capacity for adaptive deformations of state manifolds, analogous to that of fully connected networks.} \emph{what makes a quantum layer flexible enough to learn useful hidden representations, rather than merely applying rigid transformations?} We answer this by analysing how trainable quantum operations deform the manifold of quantum states at the infinitesimal level, and by extracting design rules for expressive quantum neural networks.

Concretely, we derive operational criteria that distinguish geometrically rigid ansatz structures from genuinely flexible ones, and show how these criteria can be realised through trainable data-dependent encodings with parametrised entangling directions. For practitioners, this provides a practical architecture-design checklist that reduces trial-and-error ansatz search and prioritises circuit choices with stronger representation capacity and clearer scaling prospects.

In classical learning, depth composes nonlinear maps that progressively transform an input manifold into geometries better aligned with the target task~\cite{615508,scaling_in_book,pmlr-v49-eldan16,6697897,JMLR:v21:20-345}. This geometric view helps explain why modern deep models often outperform universally approximating shallow or rigid alternatives. Quantum machine learning (QML)~\cite{Blance:2020nhl,Ngairangbam:2021yma,Abel:2022lqr,Wang:2024ygz,Fang:2024ple,Alexeev:2024dqv,Lee:2025ppl,Singh:2024dyh,Haug:2023wjv,Bal:2025ydm} models also admit universal-approximation style guarantees, but those guarantees do not, by themselves, explain which ansatz structures can realise flexible and trainable feature learning in practice. The key distinction is again adaptability: expressive models must be able to change \emph{how} they represent data, not only \emph{where} data points are moved.

The quantum setting is fundamentally constrained by symmetry. For closed systems, applying the same unitary to all states preserves unitarily invariant distances, so purely data-independent trainable unitaries are geometrically rigid. Pure data encodings do break this symmetry~\cite{Heredge:2024tua} by acting selectively on different inputs, but without trainable control, they induce fixed, non-adaptive deformations. This identifies the core requirement of QML layers: they must be both selective (data-dependent) and tunable (weight-dependent) in a coupled way.

To formalise this, we work on the pure-state manifold $\mathbb{C}P^{2^n-1}$, since physically distinct pure quantum states are rays rather than vectors, and use smooth Lie-group actions because allowed closed-system transformations are unitary and are naturally analysed through their infinitesimal generators. We introduce \emph{Classical-to-Lie-algebra} (CLA) maps that describe how classical inputs and trainable weights select directions in the Lie algebra of accessible generators. Within this framework, we identify two necessary ingredients for infinitesimal geometric flexibility: \emph{completeness}, i.e. access to all relevant generator directions through trainable parameters, and \emph{local selectivity}, i.e. state-dependent action across the data manifold. As shown below for bilinear CLA maps, exact full-rank completeness typically fails only on lower-dimensional singular sets; for this reason, the mathematically natural and practically relevant requirement is the almost-everywhere notion of \emph{almost Complete Local Selectivity} (aCLS).

Our main findings are that data-independent trainable unitaries are complete but non-selective, and therefore act as learnable rigid reorientations of the encoded quantum feature space, whereas pure data encodings are selective but non-tunable, and therefore produce fixed deformations that cannot be adaptively controlled during training. We further show that geometric flexibility requires non-trivial joint dependence on data and weights, which we characterise through CLA maps and Jacobian rank conditions (aCLS). Finally, we find that accessing the exponentially growing geometric degrees of freedom of many-qubit state manifolds requires \emph{parametrised} entangling directions; fixed entanglers such as CNOT alone cannot provide adaptive geometric control, clarifying their limited expressibility impact~\cite{Liu:2024yot}.

Taken together, these results move QNN design beyond a purely ``state-reachability'' view, in which one asks only whether a circuit family can, in principle, prepare a target state or unitary. Existing analyses are often framed in terms of controllability~\cite{Magann:2021bfp}, expressibility~\cite{Sim2019Expressibility,Larocca:2021jub,Haug:2021bca}, universal approximation~\cite{PhysRevLett.127.090506,Perez-Salinas:2021nwm,Schuld2021EffectEncoding}, or trainability diagnostics~\cite{McClean2018,Cerezo2021,Holmes2022ExpressibilityBarren,Larocca2022Diagnosing,Larocca:2024plh,Khanal:2024foz}; our viewpoint is complementary and focuses on whether layers can adaptively and selectively deform data geometry. For learning problems, this geometric criterion is essential~\cite{pmlr-v139-refinetti21b,pmlr-v267-rubin25a,Vlasic:2025hvm}: a useful model must reshape relationships between different data samples in a trainable way, i.e. it must controllably shape hidden quantum feature geometry. Additionally, within this framework, resource efficiency can constitute a \emph{quantum advantage} for classical input data, allowing one to access exponentially large real dimensions for geometric feature learning on the quantum state manifold, albeit at the cost of an exponential number of parametrised gate operations. 

Section~\ref{sec:infinitesimal_geometry} introduces the geometric classical-to-quantum bridge and develops the infinitesimal Lie-group framework, including the necessity of parametrically controllable and tunably data-dependent gates. Section~\ref{sec:geometric_definition_layers} formalises quantum layers through CLA maps and defines the almost Complete Local Selectivity (aCLS) criterion, while Section~\ref{sec:hidden_feature_geometry} applies these criteria to standard ansatz families and derives the role of parametrised entangling directions in accessing high-dimensional geometric deformations. Section~\ref{sec:numerical_examples} presents numerical illustrations of these principles, and Section~\ref{sec:summary_conclusions} closes with the main implications; technical proofs are collected in Appendix~\ref{app:proofs}.

\section{Infinitesimal geometry and quantum feature learning} 
\label{sec:infinitesimal_geometry}
\subsection{Building a geometric classical-to-quantum bridge}
Within the manifold hypothesis in classical machine learning, one treats the input domain $\X$ as an intrinsic $\dX$-dimensional manifold embedded in an ambient space $\mathbb{R}^{\din}$ with $\din> \dX$.  This means that given any point in $\X$, we can find a local region which resembles $\mathbb{R}^{\dX}$. Since this holds true for any $\mathbb{R}^n$, they are also manifolds. One of the simplest examples of a non-trivial intrinsic manifold is the 2D-sphere, whose points exist independent of how it is defined in some higher dimensions (usually 3D).  At any point on the sphere, we can move in two independent directions, and hence the local structure resembles $\mathbb{R}^2$.  

In the hidden representation of classical neural networks, it is helpful to adopt an extrinsic viewpoint, in which the manifold is mapped to $\mathbb{R}^{d}$. Each layer can continuously deform the embedded manifold, such as the 2D sphere, along any of the available $d$ real directions.  This geometric picture constitutes what is colloquially referred to as \emph{feature learning}, wherein the compositional structure distorts the initial input manifold into an appropriate, generally non-diffeomorphic geometry by tuning the weights according to the particular training objective. While one never increases the ambient dimensions of the underlying data manifold $\X$ in any of the hidden representations, the number of nodes determines the independent directions of possible continuous geometric deformations by varying the tunable weights. 

In the quantum picture, the underlying co-domain of quantum operations is no longer Euclidean and thus exhibits non-Euclidean geometry.  Specifically in the case of pure states in an $n$-qubit system, this manifold is the complex projective manifold $\mathcal{M}_n\cong\mathbb{C}P^{2^n-1}$ which takes into account the global phase ambiguity in the $2^n$-dimensional complex Hilbert space. Additionally, the possible operations on this manifold, being determined by quantum mechanics, are highly restrictive. This rather restrictive set of closed quantum systems lies within the $SU(2^n)$ unitary transformations.   While the transformations themselves are restrictive, the Lie group $SU(2^n)$, being a smooth manifold, greatly simplifies geometric analysis, as the smoothness and analyticity of the transformation on $\M_n$ allow one to infer far-reaching consequences from the induced infinitesimal geometry. This serves as the primary diagnostic tool for the analyses in this work.

To see the underlying requirement of spanning the underlying manifold, let us consider an explicit example of classifying two embeddings of $S^5$ in $\mathbb{R}^{6}$ with distinct radii $r_A$ and $r_B$, i.e. we want to identify whether samples $\x\in\mathbb{R}^{6}$ belong to $\X_0=\{|\x|=r_0 :\x\in\mathbb{R}^{6}\}$ or $\X_1=\{|x|=r_1:\x\in\mathbb{R}^{6}\}$, $r_A<r_B$. In the absence of noise, these two sets are disjoint as long as $r_0\neq r_1$; however, they are not linearly separable in $\mathbb{R}^6$.  Already, the pure state manifold $\M_2\cong\mathbb{C}P^3$ for two qubits have six real degrees of freedom.  Thus, regardless of whether one utilises a dynamic ansatz construction method for each sample, or a fixed ansatz design, a quantum algorithm which effectively discriminates $\X_0$ and $\X_1$ should span a five-dimensional subset of $\M_2$ (or any $\M_n$, $n\geq 2$ in general). More importantly, to separate $\X_0$ and $\X_1$, one can utilise the entire manifold geometry of $\M_n$ (with exponentially scaling real dimensions) to distort the quantum feature spaces to arrive at a particular region dependent on the observable that separates $\X_0$ from $\X_1$.  

Already the classical term ``\emph{hidden representation}" emphasises the notion of ``non-observability" in classical neural networks, which in the quantum case could be regarded as a physical limitation, wherein we do not explicitly measure each corresponding hidden quantum feature space. Nevertheless,  one needs to account for geometrical flexibility in such unobserved quantum features, which could then be analogously referred to as \emph{quantum feature learning}. One can formalise the notion of geometric flexibility in quantum neural networks and quantum layers in terms of the induced infinitesimal geometry. Our analysis in the remainder of the paper is therefore based on the following analogy:
\begin{quote}
As (classical) nodes in a neural network provide additional real directions for processing classical data, tunable gate parameters provide additional real directions for processing quantum states. 
\end{quote}
Specifically, we devise a concrete way to define the gate parameters' dependence on tunable weights and classical input samples, so that one can achieve infinitesimal geometric flexibility of the applied quantum operations on the initial classical data manifold $\X$.  This limit provides a concrete basis for translating the feature-learning dynamics of classical neural networks into the quantum setting.

The infinitesimal geometry of Lie group actions is captured within their corresponding Lie algebras. For our case, we work with a vector subspace of $\mf{su}(2^n)$ spanned by a subset $\mbf{G}$ of the generators $\mbf{B}=\{\sigma_x,\sigma_y,\sigma_z,I_2\}^{\otimes n}\setminus\{I_{2^n}\}$. Without the need to explicitly refer to the Lie brackets, this subspace forms the core of our analysis. Therefore, we refer to this subspace spanned by the controllable gate parameters\footnote{Note that this need not form a dynamic Lie algebra. While they are important for various aspects of state reachability and universal quantum computation diagnostics, especially for barren plateaus~\cite{Ragone:2023qbn,Fontana:2023mgx,Ragone2024} in variational algorithms, we find that the accessible $\mbf{G}$ subspace's relation to geometric flexibility presents a more basic requirement for feature learning.} as the $\mbf{G}$-subspace and regard its elements as Hermitian operators following the exponentiation convention $\exp:\mf{g}\to SU(2^n)$ as $X\mapsto e^{\frac{i}{2}\,X}$ for some $X=\sum_{i=1}^k\;\alpha_i\;G_i$ in the $\mbf{G}$ subspace. Defining the vector $\mbf{a}=(\alpha_1,\alpha_2,....,\alpha_k)\in\mathbb{R}^k$, any two states in the quantum feature space $\ket{\psi(\mbf{a}_1)}$ and $\ket{\psi(\mbf{a}_2)}$, will therefore be functions of elements $\mbf{a}_1$ and $\mbf{a}_2$ of this subspace. While there are various metric geometries that can be studied in the space of quantum states, we utilise the Fubini-Study metric due to its underlying relation between the Riemannian\footnote{See ref.~\cite{Vlasic:2025hvm} for a rigorous connection between the data manifold, the $SU(2^n)$ group manifold and thereby its induced geometry on the manifold of quantum states.} geometry on $\M_n$. The integrated Fubini-Study metric between any two states, given~\cite{PhysRevD.23.357} as  
\begin{equation} 
D_{\text{FS}}(\ket{\psi(\mbf{a}_1)},\ket{\psi(\mbf{a}_2)})=\arccos|\braket{\psi(\mbf{a}_1)}{\psi(\mbf{a}_2)}|^2=\arccos \left(F(\ket{\psi(\mbf{a}_1)},\ket{\psi(\mbf{a}_2)})\right)\quad, 
\end{equation}
\noindent 
where $F(\ket{\psi(\mbf{a}_1)},\ket{\psi(\mbf{a}_2)})=|\braket{\psi(\mbf{a}_1)}{\psi(\mbf{a}_2)}|^2$ is the fidelity between the two states. Thus, for rich feature learning, i.e., the ability to shape the underlying quantum feature space globally, one requires local geometric flexibility and control via tunable parameters.

\subsection{Necessity of parametrically controllable directions} 
\begin{table}[t] 
\begin{center} 
    \resizebox{0.8\textwidth}{!}{
    \begin{tabular}{l|c|c|c|l} 
    \toprule 
Gate-type& Data-dep. & Weight-dep. &Isometric& Geometric behaviour\\
&(Selective) & (Tunable) & & \\
\midrule 
Non-parametric & ---& ---&yes & Fixed \\
Parametric &\xmark &\cmark &yes& Learnable rigid rotation\\
Parametric &\cmark &\xmark &no& Fixed deformation\\
Parametric &\cmark &\cmark &no& Learnable deformation\\
\bottomrule
    \end{tabular}
    }
    \caption{An overview of the geometric properties of non-parametric gates and parametric gates when the parameter $\alpha$, depends on data samples and tunable weights. With non-trivial dependence on some component $x$ of the data and $w$ of the weights such that $\frac{\partial^2\alpha}{\partial w_l\partial x_m}\neq 0$, a parametrised gate has the ability to learn different geometric distortions of the underlying quantum feature space. }
    \label{tab:gate_geom} 
    \end{center} 
\end{table}

Since quantum gates are building blocks of larger circuits, we start by developing a geometric picture of group actions by single quantum gates. Given a single generator, it is straightforward to derive key global characteristics for the single-gate scenario, and we present the overall picture in Table~\ref {tab:gate_geom}.

The design of quantum circuits for QML applications generally utilises a mix of parametric single-qubit rotation gates and non-parametric 2-qubit entanglers, such as the CNOT gate, to access the non-separable parts of the Hilbert space. However, from the geometric perspective for a fixed ansatz acting on all encoded states, any non-parametric gate is fundamentally inert regardless of their entangling power since 
\begin{enumerate} 
\item  its action on any state $\ket{\psi(\mbf{a})}$ cannot be continuously controlled, thereby yielding no infinitesimal control over a state's tangent space and  
\item it can never act selectively on any two distinct states $\ket{\psi(\mbf{a}_1)}$ and $\ket{\psi(\mbf{a}_2)}$, thereby always conserving any unitarily invariant metric like $D_{FS}(\ket{\psi(\mbf{a}_1)},\ket{\psi(\mbf{a}_2)})$. 
\end{enumerate} 

Note that non-parametric gates like CNOT are important for universal quantum computation, where an arbitrary number of CNOT gates and single-qubit rotation gates can be used to dynamically construct a particular unitary transformation in $U(2^n)$. However, if the underlying VQC is fixed in the QML algorithm, the presence of non-parametric gates cannot directly improve geometric feature learning, since they are rigid by design.

\subsection{Necessity of tunable data dependence} 
Geometric rigidity is not restricted to non-parametric gates alone, since any parametric gates whose parameters are independent of input samples $\x$ can not be selective over the quantum feature space. Thus, dependence on data samples which provide selectivity is necessary for geometric deformations and quantum feature learning. Although pure-data encodings that do not have any tunable parameters break unitary invariance, they do not explicitly control the degree or nature of such breaking, which again leads to the uncontrollability of the symmetry breaking. Hence, a sufficient condition for geometric flexibility in closed  quantum systems can be formulated as:
\begin{quote}
    the ability to induce adaptive unitary symmetry breaking by utilising tunable selective unitary transformations acting on the quantum feature space. 
\end{quote}

To construct the underlying mathematical requirement for tunable selective unitary transformations, consider a parametrised gate $U(\alpha)=e^{i\;\alpha\;G}$ corresponding to some generator $G$ and real $\alpha$. Since the underlying manifold $\M_n$ is analytic, we will assume that the real parameter's dependence on weights $\w\in\W$ and input samples $\x\in\X$,  i.e. the map $\alpha:\W\times\X\to\mathbb{R}$ is analytic. For an initially encoded quantum feature space $\mc{Q}=\{\ket{\psi(\x)}=E(\x)\;\ket{\psi_0}\}$ the unitary $U(\alpha(\w,\x))$ acts on the states to give the processed quantum feature space $\mc{Q}_\w=\{\ket{\psi_\w(\x)}=e^{i\;\alpha(\w,\x)\;G}\ket{\psi(\x)}:\x\in\X\}$. The unitary operation $U(\alpha(\w,\x))$ has tunable selective unitary transformation on the feature space $\mc{Q}$ if $D_{FS}(\ket{\psi_\w(\x_1)},\ket{\psi_\w(\x_2)})$ depends explicitly on $\w$  i.e. 
\begin{equation}
    \frac{\partial{D_{FS}(\ket{\psi_\w(\x_1)},\ket{\psi_\w(\x_2)})}}{\partial w_i}\neq 0
\end{equation}
for some component $w_i$.    

For any general analytic function $\alpha(\w,\x)$, one can Taylor expand around some $\w_0$ and $\x_0$ to get 
\begin{equation} 
\begin{split} 
\alpha(\w_0+\delta\w,\x_0+\delta\x)=\alpha(\x_0,\w_0)+&\sum_{j=1}^{d_w}\delta w_j\left.\frac{\partial\alpha}{\partial w_j}\right|_{\delta w_i=0}+\sum_{i=1}^{\din}\delta x_i\left.\frac{\partial\alpha}{\partial x_i}\right|_{\delta x_i=0}\\
&+\sum_{i,j}\delta w_j\;\delta x_j\left.\frac{\partial^2\alpha}{\partial w_i\partial x_j}\right|_{\delta w_j=0,\delta x_i=0}+...
\end{split} 
\end{equation}
We can analyse the geometric role of the first-order terms as follows: 
\begin{itemize} 
\item for data-independent unitary gate, $\left.\frac{\partial\alpha}{\partial x_i}\right|_{\delta x_i=0}=0$ for all components of $\x,$ which leads to fixed deformations of the entire quantum feature space.  
\item for pure data-encoding unitary gate, $\left.\frac{\partial\alpha}{\partial w_j}\right|_{\delta w_j=0}=0$ for all components of $\w$, which leads to non-adaptive deformations of the entire quantum feature space.  
\end{itemize} 
Thus, a necessary condition for tunable selective unitary operation is that $\left.\frac{\partial\alpha}{\partial x_i}\right|_{\delta x_i=0}\neq0$ and $\left.\frac{\partial\alpha}{\partial w_i}\right|_{\delta w_i=0}\neq0$ for some components $x_i$ and $w_j$. However, this alone is not sufficient since $\alpha(\w,\w)=f(\x)+g(\w)$ satisfies this condition but the unitary gate operations factorises as $U(\w,\x)=e^{i\;f(\x)\;G}\;e^{i\;g(\w)\;G}$ leading to  $\frac{\partial{D_{FS}(\ket{\psi_\w(\x_1)},\ket{\psi_\w(\x_2)})}}{\partial w_i}= 0$. A sufficient condition for tunable selective unitary gate operation is, therefore, non-zero mixed partial derivatives $\frac{\partial^2\alpha}{\partial w_j\partial x_i}\neq0$ with additional spectral requirements on $G$, which is captured in the following:

\begin{proposition}
\label{prop:gate_tun_sel_act}
    A parametrised quantum gate $e^{i\;\alpha(\w,\x)\;G}$ acting on $\mc{Q}$  has tunable unitary action under the weight $w_j$ i.e. $\frac{\partial{D_{FS}(\ket{\psi_\w(\x_1)},\ket{\psi_\w(\x_2)})}}{\partial w_j}\neq 0$ if $\frac{\partial^2\alpha}{\partial w_j\partial x_l}\neq 0$ and $G$ has at least two distinct eigenvalues that are not orthogonal to the states $\ket{\psi(\x_1)}$ and $\ket{\psi(\x_2)}$. 
\end{proposition}
\noindent 
The proposition is proved in Appendix~\ref{sec:gate_tun_sel_act}. 

Another important criterion for geometric flexibility beyond explicit dependence is to have non-monotonic dependence on $w_j$ such that different values of $w_j$ can increase or decrease the distance between two encoded states. This is ensured by $D_{FS}$ having both negative and positive valued derivative values. Since this behaviour is entirely dependent on the fidelity (of which $D_{FS}$ is an implicit function), from Eq~\ref{eq:fid_diff_weight}, we can write the weight-dependent parts of the derivative as  

\begin{equation*} 
\begin{split} 
    \frac{\partial F}{\partial w_j}\propto& \frac{\partial\Delta \alpha}{\partial w_j}\;e^{i\Delta\alpha\;(\lambda_a-\lambda_b)}\quad,
    \end{split}
\end{equation*}
where $\Delta \alpha=\alpha(\w,\x_1)-\alpha(\w,\x_2)$, and for simplicity, we have assumed that $\lambda_a$ and $\lambda_b$ are the only two non-degenerate eigenvalues of $G$. Thus, for any two generic states $\ket{\psi_\w(\x_1)}$ and $\ket{\psi_\w(\x_2)}$, the fidelity's sign can be controlled through the interplay of $\frac{\partial\Delta \alpha}{\partial w_j}$ or  $e^{i\Delta\alpha\;(\lambda_a-\lambda_b)}$. To leave the learning dynamics primarily in the quantum domain, we will take $\alpha(\w,\x)$ to be bilinear in $\w$ and $\x$, thus making the term $\frac{\partial\Delta \alpha}{\partial w_j}$ independent of $w_j$. In this case, the sign is primarily controlled by the phase factor, which still allows the derivative to be positive or negative depending on the weights' values. 

\begin{figure}[t]
\centering 
\includegraphics[width=0.325\textwidth]{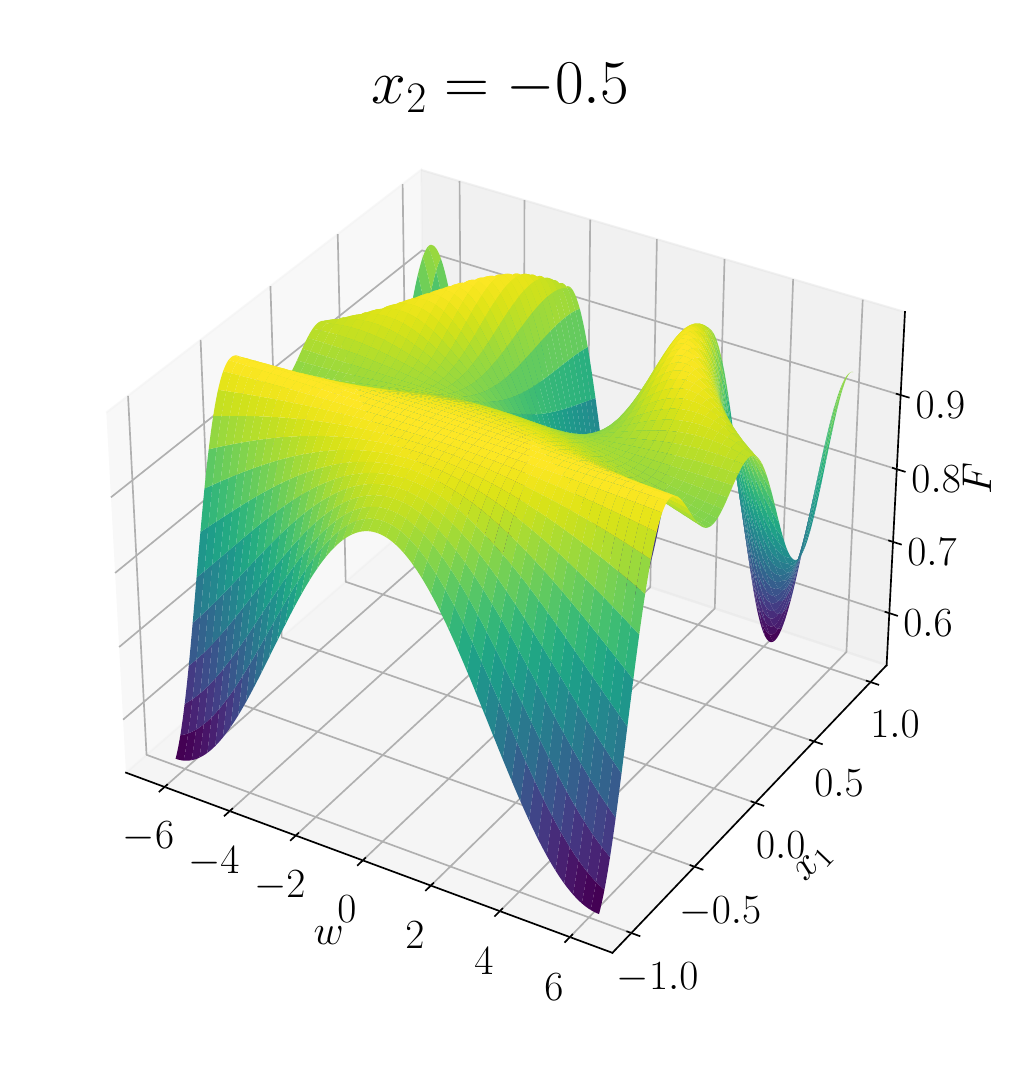}
\includegraphics[width=0.325\textwidth]{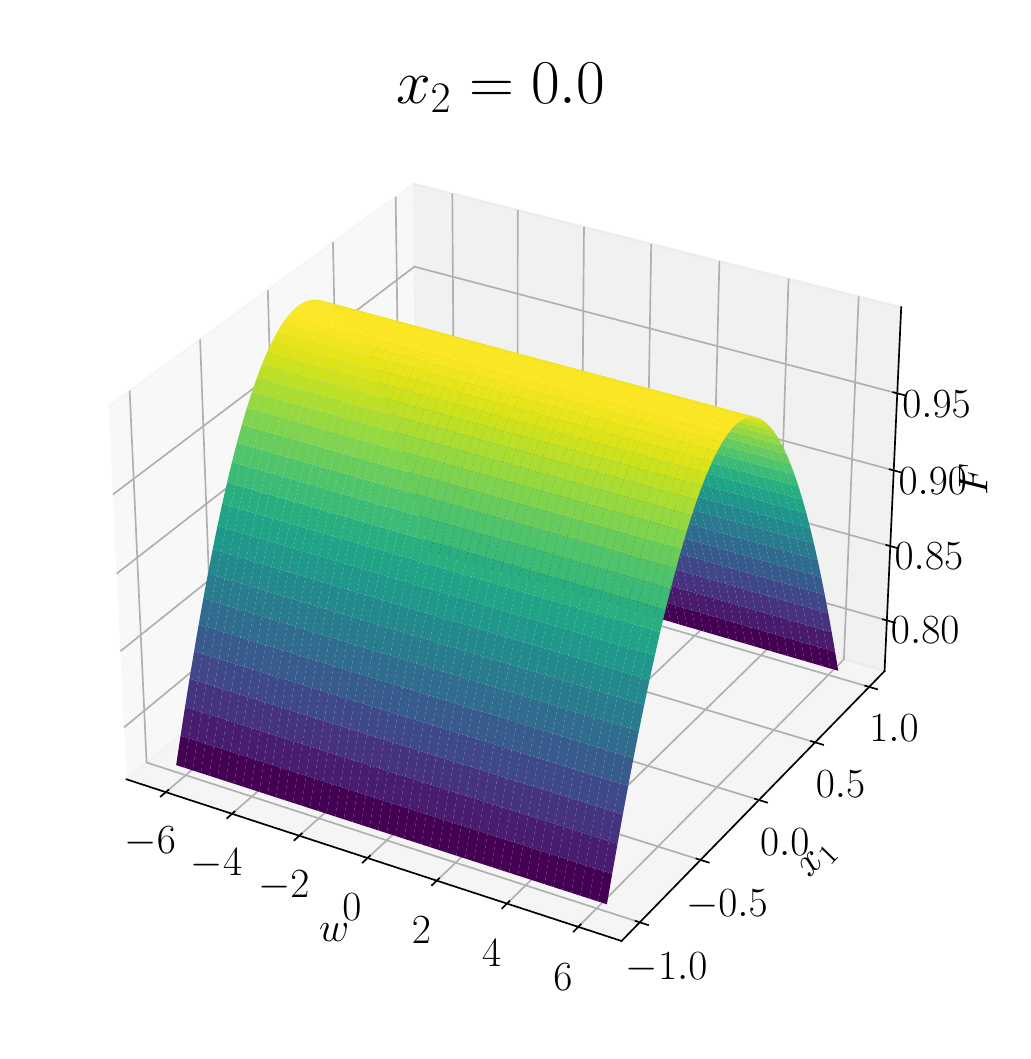}
\includegraphics[width=0.325\textwidth]{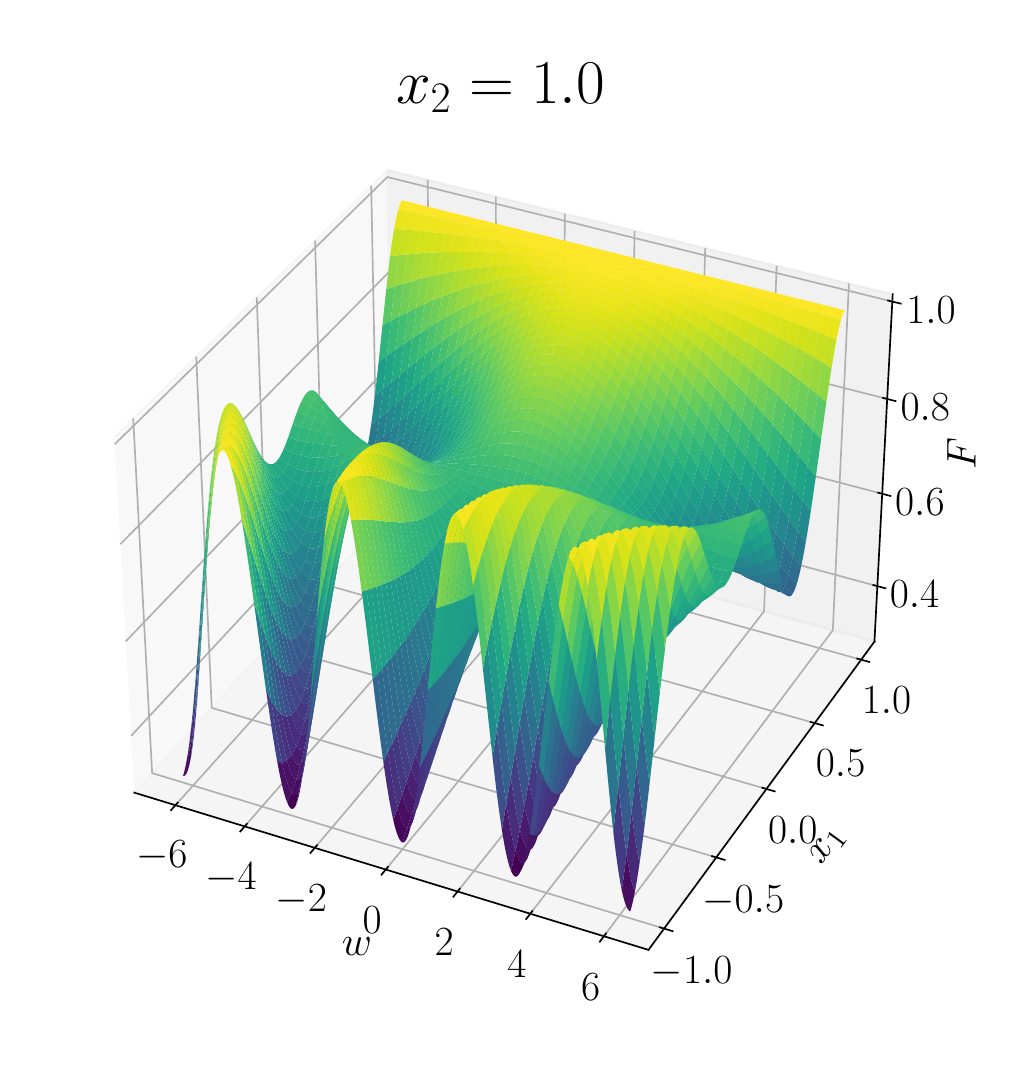}\\
\includegraphics[width=0.325\textwidth]{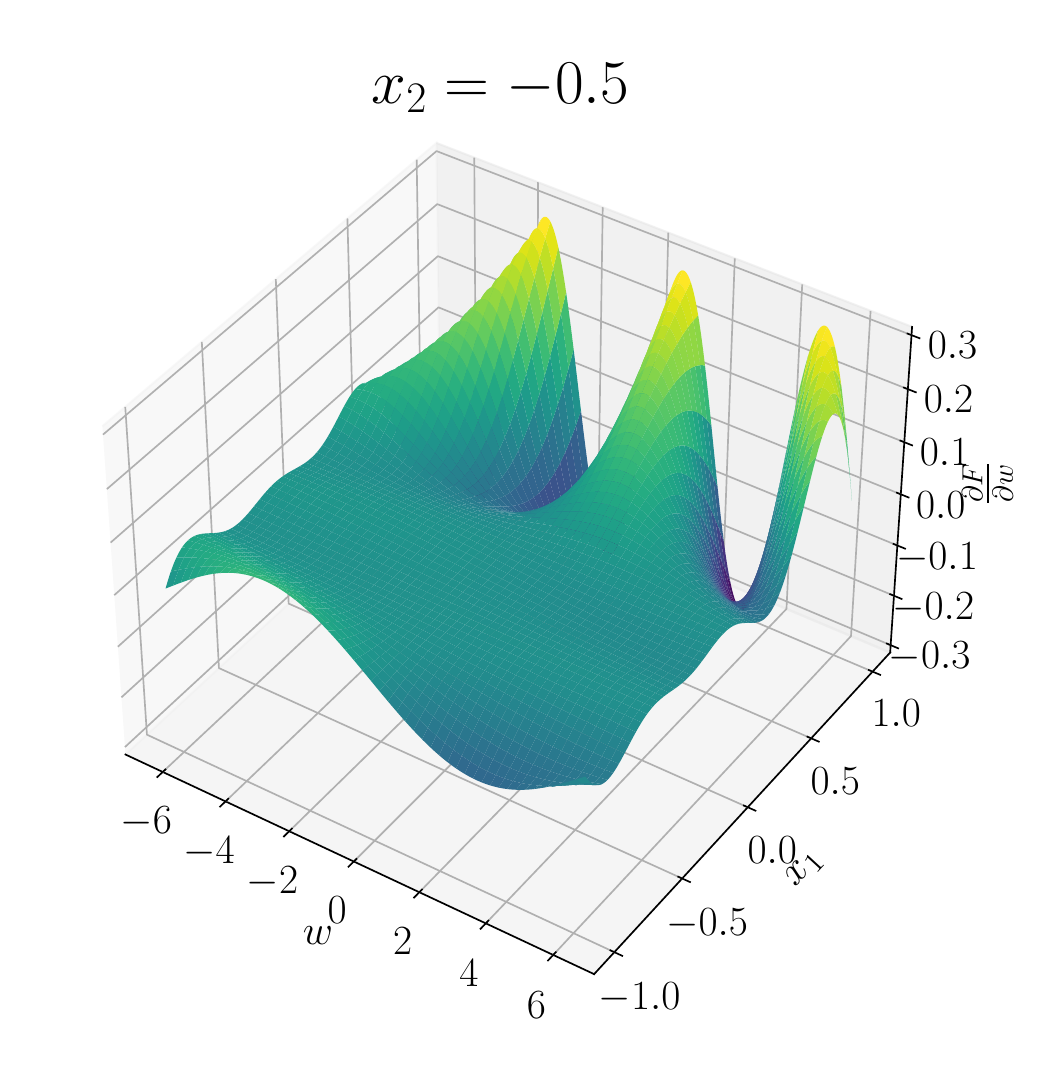}
\includegraphics[width=0.325\textwidth]{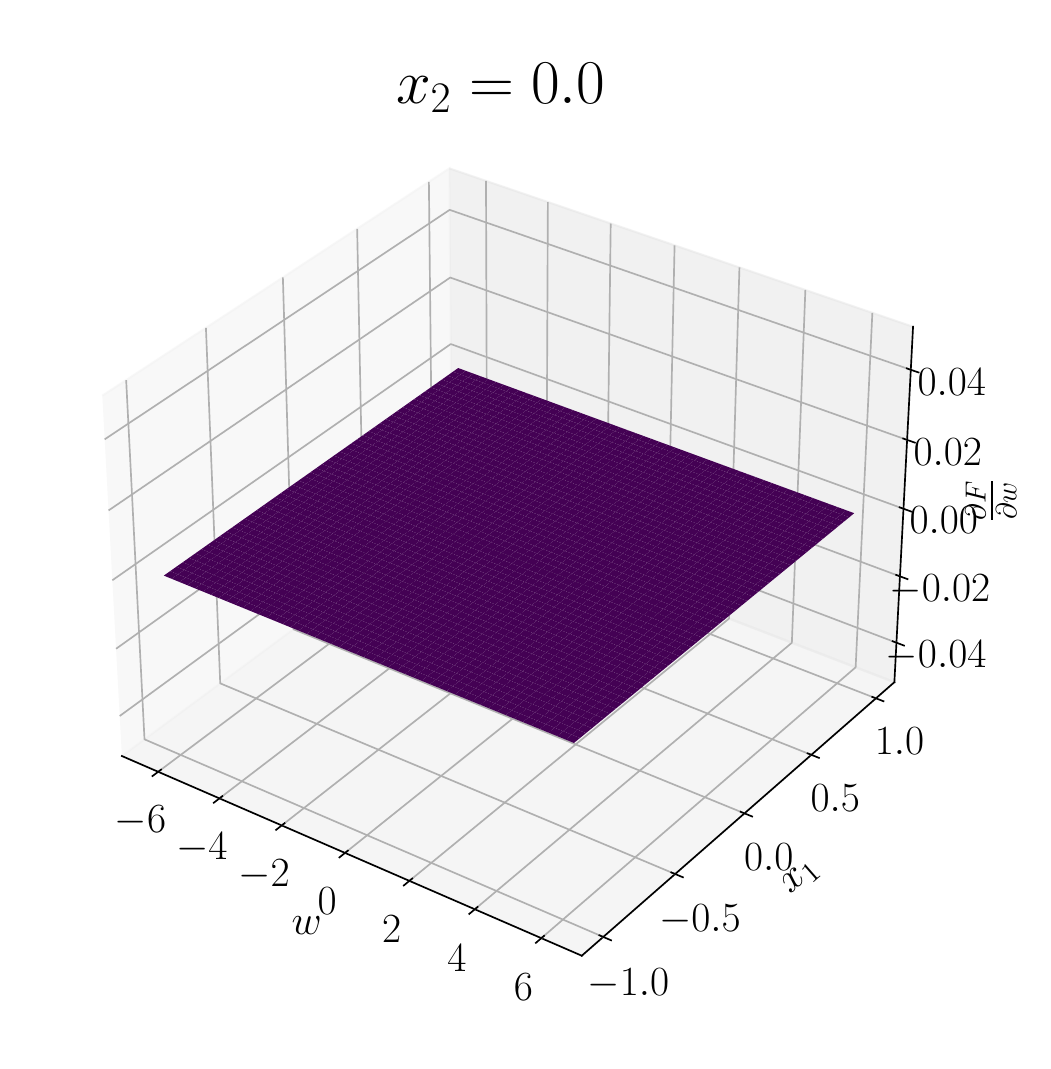}
\includegraphics[width=0.325\textwidth]{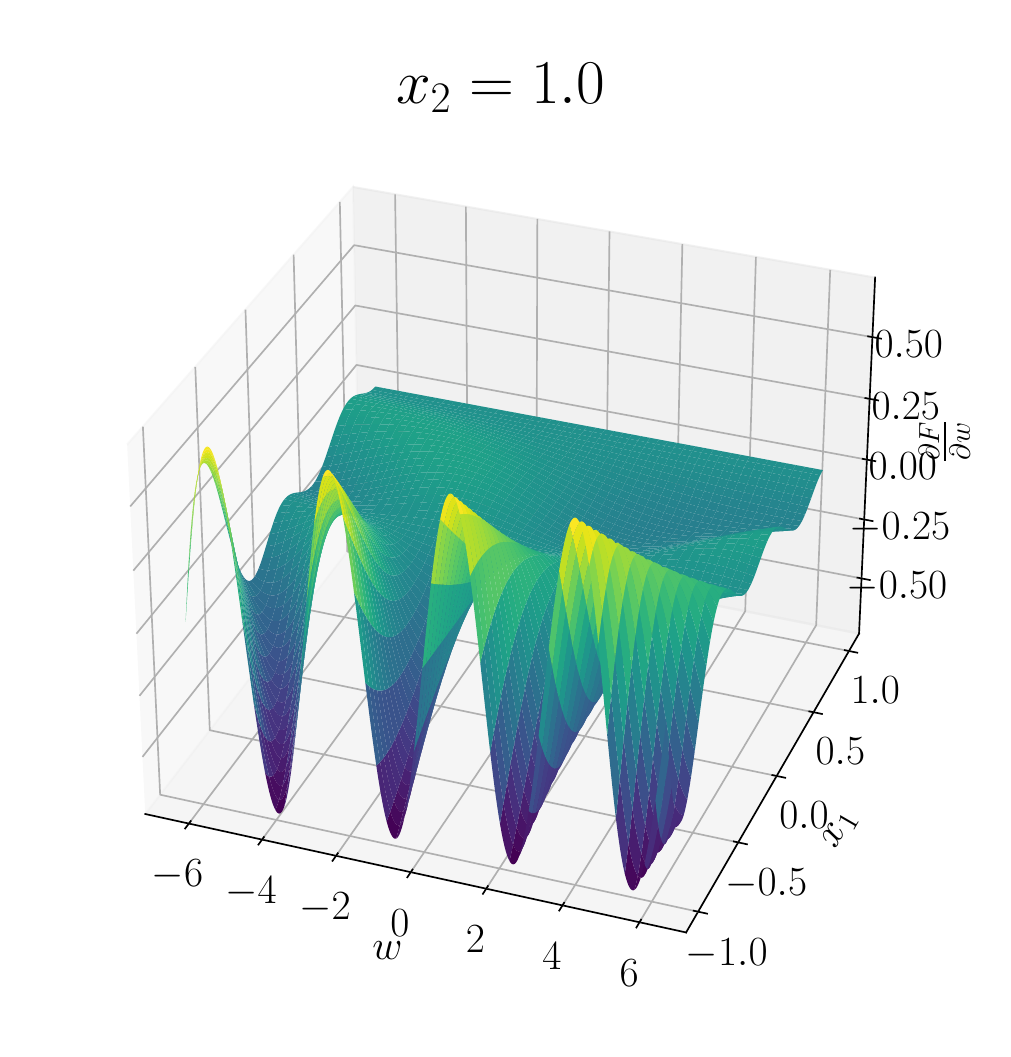}
\caption{Fidelity (top) between two states $\ket{\psi_w(x_1)}$ and $\ket{\psi_w(x_2)}$ where $\ket{\psi_w(x)}=e^{\frac{i}{2}wx\sigma_z}\;e^{\frac{i}{2}x\sigma_x}\ket{0}$. The partial derivative $\frac{\partial F}{\partial w}$ with respect to the tunable parameter $w$ is shown in the bottom row. We see that when $x_2=0$, since $\ket{\psi_w(x_2)}=\ket{0}$, fidelity is independent of $w$ with a zero partial derivation for any value of $x_1$.  
}
\label{fig:fid_plots} 
\end{figure}

In appendix~\ref{sec:fid_1d_domain}, we work out the fidelity of a 1D domain encoded via a Pauli-X rotation gate and rotated with bilinear input-dependent Pauli-Z rotations that  satisfy the sufficient condition. Taking some specific values of $x_2$, we show the variation of fidelity and its partial derivative with respect to $w$ for different values of $x_1$ in figure~\ref{fig:fid_plots}.  Explicitly, we see that when $x_2=0$, the initial state $\ket{0}$ being an eigenvalue of $\sigma_z$ loses the ability to effectively tune the distance shown as zero partial derivatives over all values of $w$ and $x_1$.

\section{Towards a geometric definition of quantum layers}
\label{sec:geometric_definition_layers}
Although we have primarily analysed the geometry of a single parametrised gate's action, one already sees a concrete connection between the infinitesimal geometry on $\M_n$ induced by the unitary group action and the overall geometry of the quantum feature space. This is a specific feature of the Riemannian structure of the manifold $\M_n$ where the global metric $D_{FS}$ is induced by the infinitesimal Riemannian metric tensor $g_{ij}$. Additionally, since changes in $\M_n$ are governed by unitary transformations through its action on states, the available directions in the group manifold decide the underlying geometric changes in $\M_n$. 
Therefore, we will study the infinitesimal geometric structure induced by a parametrised circuit composed of different generators as a means to construct quantum layers that can learn geometrically rich features within each hidden quantum representation. 

\subsection{Classical-to-Lie-algebra maps}
Given a circuit, let $\mbf{G}=\{G_1,G_2,...,G_k\}$ denote the set of unique generators corresponding to each  parametrised gate through the unitary transformation
\begin{equation}
\label{eq:unit_wx}
U(\mbf{w},\x)=\prod_{j=1}^k\;e^{\frac{i}{2}\,\alpha_j(\w_j,\x)\,G_j}\quad,
\end{equation} 
in some particular order. $\alpha_j:\W\times\X\to\mathbb{R}$ are predefined maps with independent weights $\w_j$, and $\w=(\w_1,\w_2,...,\w_k)\in\mathbb{R}^{d_w\times k}$ is the entire weight vector for the set of generators $\mbf{G}$. Expanding the exponentials,  we get $$U(\w,\x)=\left(1+\frac{i}{2}\sum_{j=1}^k\,\alpha_j(\w_j,\x)\,G_j+O(\alpha^2)\right)$$
Hence, to study the infinitesimal geometric changes induced by a block of parametrised quantum gates, one can study the action of the subspace spanned by the linear term. Therefore, we define a Classical-to-Lie-Algebra (CLA) map as follows:
\begin{definition} A classical-to-Lie-algebra (CLA) map is defined as a bilinear map $\mathbf{a}:\W\times\X\to\mathbb{R}^{k}$ with components denoted as $\alpha_j(\w,\x)$ via which we map to elements of the Lie algebra $\mf{g}$ as 
\begin{equation}
    \Gamma(\w,\x)=\sum_{j=1}^k\,\alpha_j(\w_j,\x)\;G_j\quad.
\end{equation}
\end{definition}
\noindent 
Note that we define the map to be bilinear since we are primarily interested in infinitesimal behaviour---namely, how the extrinsic geometry of the image of the quantum operations acting on $\X$ varies for local neighbourhoods in the manifold and how these can be changed by varying the weights $\w$. Moreover, bilinearity satisfies the sufficient condition of Proposition~\ref{prop:gate_tun_sel_act} for tunable selective unitary action and unifies data-independent and fixed-data encoding ansatzes as trivial dependencies on the absent factors.  It also facilitates simple classical-to-quantum maps in QML models, with task-specific nonlinearities emerging from the underlying manifold geometry.

To study dependence on both weights and data, we define  maps at fixed weights $\mathbf{a}_\w:\X\to\mathbb{R}^k$, and fixed input samples $\mathbf{a}_\x:\mathbb{R}^{d_w\times k}\to\mathbb{R}^k$ with the assignment
\begin{equation}
\begin{split}
    \x\mapsto \mathbf{a}(\w,\x)\quad,\quad\w\mapsto \mathbf{a}(\w,\x)\quad.\\
    \end{split} 
\end{equation}
We will identify two key attributes of each, in isolation (i.e., completeness and local selectivity), pointing to the need to unify both characteristics for a richer geometry. Since we have assumed $\mbf{a}(\w,\x)$ to be bilinear, these are linear in each of their arguments.

\subsection{General structure of quantum layers}
Before constructing geometrically flexible CLA maps, we contextualise them within data-reuploading quantum circuits and clarify each factor's intuitive geometric role. We define an $l^{th}$-quantum layer acting on quantum states as a variational quantum ansatz of the form 
\begin{equation}
\label{eq:q_layer} 
    \ket{\psi^{(l)}_{\bar{\w}_{l}}(\x)}=U_l(\w_l,\x)\;\ket{\psi^{(l-1)}_{\bar{\w}_{l-1}}(\x)}
\end{equation}
where $U_l(\w_l,\x)$ corresponds to the circuit's unitary transformation. Thus, through its corresponding CLA map  $\mbf{a}^{(l)}:\mathbb{R}^{k_l\times d_{w_l}}\times\X\to\mathbb{R}^{k_l}$, we map to the $\mbf{G}_l$-subspace as  $$\Gamma_l(\w_l,\x)=\sum_{j=1}^{k_l}\;\alpha_j^{(l)}(\w_j^{(l)},\x)\,G^{(l)}_j\quad.$$
To accommodate for both pure data-encodings and data-independent tunable unitaries, we have left the generators arbitrary in each layer and defined $\bar{\w}_{l}=\oplus_{a=1}^{l}\w^{(a)}$ and $\ket{\psi^{(0)}_{\bar{\w}_{0}}(\x)}=\ket{\psi_0}$.  
Thus, we can define hidden quantum feature spaces $\mc{Q}^{(l)}_{\bar{\w_{l}}}$ as the range of the states $ \ket{\psi^{(l)}_{\bar{\w}_l}(\x)}$ in $\M_n$ for some parameter $\bar{\w}_l$ and input manifold $\X$. Moreover, we are now ready to study the geometric changes that can be induced from $\mc{Q}^{(l)}_{\bar{\w}_{l}}$ to $\mc{Q}^{(l+1)}_{\bar{\w}_{l+1}}$ under infinitesimal changes in the weights in local neighbourhoods in $\X$.

In the first layer $l=1$, acting on some reference state $\ket{\psi^{(0)}(\x)}\equiv\ket{\psi_0}$, the quantum feature space $\mc{Q}^{(1)}_{\bar{\w}_{1}}$  is the range of all states 
\begin{equation*}
\ket{\psi^{(1)}_{\bar{\w}_1}(\x)}=U_1(\w_1,\x)\ket{\psi_0}
\end{equation*}
at some fixed $\w_1$ for all $\x \in\X$. Depending on the exact value of $\w$ and the structure of $U_1(\w_1,\x)$, this may or may not be a strict \emph{manifold embedding}.\footnote{Similarly, the input layer in FCNNs is generally not invertible as the processing already starts from the first layer, where different weights lead to non-isometric geometries.} Within this smooth map, the local structure of how the geometry of the manifold $\X$ changes is captured by the differential map induced in $\mc{Q}^{(1)}_{\w_1}(\x)$, between the tangent space at $\x$ in $\X$ to the tangent space of $\ket{\psi^{(1)}_{\bar{\w}_1}(\x)}$ in $\M_n$. This geometric change in $\M_n$ is determined by the CLA map dependent only on the initial state $\ket{\psi_0}$,  wherein the quantum feature space $\mc{Q}^{(1)}_{\bar{\w}_{1}}$ is the set of all states traversed by family of unitary operations $U_1(\w_1,\x)$ over $\x\in\X$ acting on $\ket{\psi_0}$.  

From the second layer onward, the state depends nontrivially on the input sample.  To highlight the importance of the CLA maps in deciding the geometry at any layer $l$, we can write 
\begin{equation}
\label{eq:qnn_expand}
    \ket{\psi^{(l)}_{\bar{\w}_{l}}(\x)}=U_l(\w_l,\x)\,U_{l-1}(\w_{l-1},\x)...U_1(\w_1,\x)\;\ket{\psi_0}\quad. 
\end{equation}
Thus, for fixed $\bar{\w}_l$, and a particular input domain $\X$, the geometry of the underlying quantum feature space $\mc{Q}^{(l)}_{\bar{\w}_{l}}$ is completely determined by the behaviour of the unitary transformations, the infinitesimal structure of which is determined by their respective CLA maps. 

Physically, the $\mathbf{G}_l$-subspace of $\mathfrak{su}(2^n)$ determines the accessible directions for quantum state manipulation. By making elements of this subspace data dependent (via the manifold $\X$), we break the underlying unitary geometry in $\M_n$, inducing non-isometric geometries. Each application of data-dependent unitaries maps the data manifold into the group manifold of $SU(2^n)$, acting selectively on states in $\mc{Q}^{(l)}_{\bar{\w}_{l}}$ deforming the geometry to give $\mc{Q}^{(l+1)}_{\bar{\w}_{l+1}}$. Locally, the structure of the encoding $\X\mapsto SU(2^n)$ is captured by the CLA map, thereby determining the geometric flexibility of the algorithm. For any $l>1$, $\mc{Q}^{(l)}_{\bar{\w}_{l}}\subset\M$ describes \emph{hidden quantum feature spaces}, and we diagnose conditions on the CLA map that enable rich feature learning capabilities.  

\subsection{Almost Complete Local Selectivity} 
To study the range of geometries induced by CLA maps, we drop the explicit layer index $l$ and define 
\begin{equation}
\label{eq:sing_layer}
\ket{\psi'_\w(\x)}\equiv U(\w,\x)\ket{\psi(\x)}
\end{equation} focussing on geometric changes within a single operation from the feature space $\mc{Q}\ni\ket{\psi(\x)}$ to $\mc{Q}_\w\ni\ket{\psi'_\w(\x)}$. 
Taking its partial derivative with some component $w_{j,i}$ of $\w_j$, and expanding the exponentiated unitary to first order in $\alpha_j$ we have
\begin{equation*}
  \frac{\partial\ket{\psi'_{{\w}}(\x)}}{\partial w_{j,i}}\approx \;\frac{\partial \alpha_{j}}{\partial w_{j,i}}\;G_j\;\ket{\psi(\x)}
\end{equation*} 
This expression captures the movement of the state $\ket{\psi(\x)}$ in the manifold $\M_n$ in the direction acted upon by the generator $G_j$ for infinitesimal changes in the parameter $w_{j,i}$. 

For maximum infinitesimal geometric flexibility, varying the weights should be able to locally access all directions in the $\mbf{G}$-subspace for any given $\x\in\X$. At a particular $\x\in\X$, we call this property \emph{completeness} wherein the map from the weight space $\mc{W}=\mathbb{R}^{d_w\times k}$ is locally surjective to the $\mbf{G}$-subspace. Writing the weight components $w_{j,i}$ with single indices $(\w_1,\w_2,...,\w_k)=(\omega_1,\omega_2,...,\omega_{d_w\times k})\in\mc{W}$, the CLA map is locally surjective for some fixed $\x$,  when the weight Jacobian matrix $\mbf{J}_{_{{\mc{W}}}}(\x)$ with components $\frac{\partial \alpha_j}{\partial \omega_{i}}$ has rank $k$. While easily satisfied for data-independent CLA maps, the Jacobian's linear $\mathbf{x}$-dependence in non-trivially bilinear maps prevents the rank condition from holding generally across $\mathcal{X}$, failing on zero-volume sets. Hence, over the entire $\X$, we consider \emph{almost complete} CLA maps where the rank condition can fail in subsets of measure zero. 

If the CLA map is independent of $\x$, the corresponding transformation simply reorients the local neighbourhood without changing its unitary geometry in $\M_n$. Selectivity in local neighbourhoods of $\ket{\psi(\mathbf{x})}$ is thus necessary for flexible global geometries. The local structure of the unitary transformation 
$$\frac{\partial U(\w,\x)}{\partial x_i}\approx \sum_j\frac{\partial\alpha_j}{\partial x_i}\,G^{(l)}_j\quad$$
determines the range of transformations in the group manifold $SU(2^n)$ that can be selectively applied to the state $\ket{\psi(\x)}$. Clearly, a locally injective map $\mbf{a}_\w:\X\to\mathbb{R}^k$ utilises the full span of selectivity possible in the neighbourhood of some point $\x\in\X$. On the other hand, injectivity for any $\w$ leads to unnecessary and excessive application of selective unitary transformations that attempt to distort every local region in $\mc{Q}$.    
Since this local injectivity condition is determined by the data Jacobian matrix $\mbf{J}_{\X}=\left[\frac{\partial\alpha_j}{\partial x_i}\right]$ (having maximal rank of $\dim(\X)=d_\X$), a weight dependent $\mbf{J}_\X(\w)$ for which the maximal rank is achievable in some non-negligible region of the weight space indicates non-rigid mapping of the local neighbourhood to the $\mbf{G}$-subspace. For non-trivial bilinear dependence, we call this property \emph{local selectivity}. Combining the two properties, we can formally define the almost Complete Local Selectivity (aCLS) property of CLA maps as follows.
\begin{definition}
A classical-to-Lie-algebra map to the underlying $k\leq 4^n-1$ dimensional $\mbf{G}$-subspace of $\mf{su}(2^n)$ is almost complete if the weight Jacobian matrix $\mbf{J}_\mc{W}(\x)=\left[\frac{\partial\alpha_j}{\partial\omega_i}\right]$ has rank $k$ for almost every $\x\in\X$. It is locally selective if the data Jacobian matrix $\mbf{J}_\mc{X}(\w)=\left[\frac{\partial\alpha_j}{\partial x_i}\right]$ has rank $\dim(\X)$ for a non-negligible set of weights $\w\in\mc{W}$.
\end{definition}

\section{Geometry of hidden quantum feature spaces} 
\label{sec:hidden_feature_geometry}
Having discussed the rank conditions on the Jacobian matrices for the aCLS property to hold, we can now diagnose the geometric properties of common ansatz structures and construct CLA maps with rich feature-learning capabilities.

\subsection{Global non-selectivity of data-independent unitaries}
Many VQC ansatz have an initial data-encoding, say $U_1(\w_1,\x)\equiv E(\x)$ with $d=k_e$
\begin{equation}
    \Gamma_1(\x)=\sum_{j=1}^{k_e} \;x_j\;  G_j
\end{equation}
with subsequent ``layers" forming a block of pure-tunable unitaries, say $U_l(\w_l,\x)\equiv S_l(\w_l)$ with the CLA map 
$$\Gamma_l(\w_l)=\sum_{j=1}^{k}\;w^{(l)}_j\;K^{(l)}_j\quad.$$
Here, we define as $\mbf{G}_1\equiv\{G_1,G_2,...,G_d\}$ and $\mbf{G}_l\equiv\mbf{K}_l=\{K^{(l)}_1,K^{(l)}_2,...,K^{(l)}_k\}$ for $l>1$.  

The hidden quantum feature spaces under such a setup are not only diffeomorphic to the initially embedded input quantum feature space but are \emph{rigid isometries}. Under the aCLS property, this happens because the data Jacobian matrix $\mbf{J}^{(l)}_\X$ is the zero-matrix for all $l>1$, and is therefore non-selective everywhere on $\X$. Any variational ansatz in which the first layer encodes the data and subsequent layers are data-independent tunable unitaries lacks a geometric notion of  ``depth" in the sense of classical neural networks, and the geometry remains infinitesimally and globally rigid. Such quantum models are closer to kernel methods~\cite{Havlicek:2018nqz,Schuld:2018uel,Schuld:2021hzr,Kempkes:2025hiw} than to neural networks and the entire unitary transformation after the initial encoding serves to find an orientation of the initial quantum feature space with respect to the eigenvectors of observables that best approximate the target function. 

Even within the isometric mappings provided by sequential data-independent unitaries, if each $S_l$ is identically  characterised by a repeating circuit block with generator set $\mbf{K}\equiv\mbf{K}_2=\mbf{K}_3...=\mbf{K}_L$, we get the resultant unitary transformation, say $\bar{S}(\w)$ as  
\begin{equation*}
    \bar{S}(\w)=\prod_{l=1}^L\;S(\w_l) \quad, 
\end{equation*}
$\w=(\w_1,\w_2,...,\w_L)$. 
Therefore, the resulting $\mbf{G}$-subspace of the unitary operator $\bar{S}(\w)$ is at most $k$ dimensional and locally has the CLA map
\begin{equation*}
\bar{\Gamma}(\w)=\sum_{j=1}^{k} \left(\sum_{l=2}^L\alpha^{(l)}_j(\w_l)\right)\; K_j
\end{equation*} 
whose rank cannot exceed $k$.  This exposes a critical geometric limitation of repeated layers in the absence of data re-uploads: the effective local rank of the entire unitary circuit is fundamentally limited by the number of distinct generators corresponding to parametrically tunable quantum gate operations. More generally, given a data-independent unitary ansatz $U(\w)$ consisting of a finite number of gates acting on encoded states $\ket{\psi(\x)}$, the maximum rank of the weight Jacobian matrix of the corresponding CLA map is limited by the number of unique generators of parametrically controllable gates which forms the dimensions of the $\mbf{G}$-subspace. Explicit examples for such rigid rank conditions are given in Appendix~\ref{sec:rigid_isom}.

\subsection{Non-adaptive selectivity in pure-data re-uploading}

Data-reuploads~\cite{PerezSalinas2020DataReuploading} constitute one of the ways to make quantum models more expressive and are known to be asymptotically dense in $L_2(\X)$ by accessing higher Fourier decompositions~\cite{Perez-Salinas:2021nwm,Schuld2021EffectEncoding}. However, the commonly utilised ansatz, in which a repeating block of pure-data encodings is interspersed with data-independent, tunable unitaries, lacks infinitesimal geometric flexibility due to an unbalanced trade-off between completeness and local selectivity. 

Assuming that the data encoding and tunable unitaries consist of two different circuits characterised by generator sets $\mbf{G}=\{G_1,G_2,...,G_k\}$ and $\mbf{K}=\{K_1,K_2,...,K_t\}$, respectively, we can write down the unitary transformations as 
\begin{equation}
    E(\x)=\prod_{j=1}^k\;e^{\frac{i}{2}\,\bar{x}_j\,G_j}\quad,\quad S(\w)=\prod_{j=1}^t\;e^{\frac{i}{2}\,w_j\,K_j}
\end{equation}
where we have defined $\mbf{\bar{x}}\equiv(\bar{x}_1,\bar{x}_2,...,\bar{x}_k)$ formed by repeating the input $k/d$ times (implicitly considering that $k$ is a multiple of $d$). 

Repeating this multiple times in the ansatz, we can write down the general structure of the CLA map for $l\geq 1$  as 
\begin{equation}
\begin{split} 
    \Gamma_{2l-1}(\x)&=\sum_{j=1}^{k} \;\bar{x}_j\;  G_j\\
    \Gamma_{2l}(\w_{2l})&=\sum_{j=1}^{t}\;w^{(2l)}_j\;K_j\quad.
    \end{split} 
\end{equation}
Thus, one can dissect the CLS property of such maps as follows: 
\begin{itemize}
\item pure data-encoding CLA maps ($\Gamma_{2l-1}$) are absolutely locally selective since $\text{rank}(\mbf{J}_\X^{(2l-1)})=\dim(\X)$ but are directionally rigid with  $\text{rank}(\mbf{J}_\mc{W}^{(2l-1)})=0$
\item data-independent CLA maps ($\Gamma_{2l}$) are globally non-selective $\text{rank}(\mbf{J}_\X^{(2l)})=0$ but is complete with  $\text{rank}(\mbf{J}_\mc{W}^{(2l)})=k$
\end{itemize} 

In the context of the CLS property, each data-encoding unitary maps a particular element of the domain $\X$ to a fixed element of the group $SU(2^n)$. This corresponding transformation remains the same across all layers, since the weight Jacobian rank of the pure-data-encoding CLA maps is zero. While this could, in principle, be sidestepped by flexible data-independent unitaries that orient the initial quantum feature space to arbitrary orientations per layer,\footnote{In fact, this is the central mechanism relied upon in proofs of universality.} the non-rigid geometric changes induced by the network remain non-tunable and thus have limited practical utility for feature learning. 
Thus, such quantum layers can orient the input manifold and distort it in fixed directions, but cannot parametrically control the degree of these distortions.

\subsection{Geometrically flexible CLA maps} 
From the discussions above, we see that for a rich array of geometric deformations, we want two properties together: the ability to steer the CLA map into any direction in the $\mbf{G}$-subspace by continuous parameter dependence (completeness) and for this change to be possibly selective at different locations in the quantum feature space. The reason pure data-encoding and data-independent CLA maps fail to satisfy the CLS property is due to the requirement of non-trivial dependence of $\mbf{a}(\w,\x)$ on both weights and the input samples to simultaneously make the two Jacobian matrices of sufficient rank.

As already remarked above, completeness in the entire domain $\X$ is rather strong; we will explicitly construct bilinear CLA maps and see why this occurs. First, consider a simple case where $d\equiv \din=k$ and define
\begin{equation*}
    \alpha_j(w_j,x_j)=w_j\,x_j\quad,
\end{equation*}
from which the Jacobian matrices evaluate to $\mbf{J}_\mc{W}(\x)=\text{diag}(x_1,x_2,...,x_d)$ and $\mbf{J}_\X(\w)=\text{diag}(w_1,w_2,...,w_d)$. It is straightforward to see that the map is locally selective at any fixed $\x$ since the data Jacobian $\mbf{J}_\X(\w)$ will have rank $d$ for some $\w\in\mathbb{R}^d$ that does not have non-zero elements. For completeness, we require $\text{rank}(\mbf{J}_\mc{W})$ to be $d$ in the entire $\X$. While this will be true for most values of $\x\in\X$, there will be a drop in rank whenever there are zero components in $\x$. Thus, the CLA map will not be complete. However, since the set of points with at least one zero component is strictly lower dimensional, it remains almost everywhere complete as the Lebesgue measure of such singular sets is zero.\footnote{For intrinsic manifolds with $\dim(\X)<d$, one has to evaluate the $\dim(\X)$-dimensional measure in $\mathbb{R}^d$ under which the set comprising of all $\x$ where the condition does not hold will contribute to zero measure since they will be lower dimensional.}

While such a simple example suffices for presenting the difficulty in constructing bilinear and complete CLA maps, the local selectivity endowed in each generator direction by the $\x$ dependence is limited since it depends on a single constituent of $\x$. Additionally, in the general case, the generators $G_i$ may not act in a similar fashion on $\M_n$, and may be a mixture of any $m$-gate operations. We will explicitly discuss such aspects and the need for parametrised entangling gates to access the exponentially high real-dimensional scaling of $\M_n$ with the number of qubits $n$ in the next subsection. To have tunable local selectivity dependent on each component of the input component for a single generator direction, we can choose $d_w=d $, construct weight vectors $\w_j\in\mathbb{R}^d$ for each gate $G_j$ to define components of the CLA map as 
\begin{equation}
\label{eq:aCLS_comp}
    \alpha_j(\w_j,\x)= \w_j\cdot\x \quad. 
\end{equation}
The redundant weights in each generator direction additionally help in reducing the (already negligible) set of points where the local selectivity fails when $\x=\mbf{0}$ since the weight Jacobian matrix becomes block diagonal constructed out of $d\times d$ blocks each having the vector $\x$ as its diagonal elements. Note that there is still local selectivity around $\mbf{x}=\mbf{0}$, since any other point in its local neighbourhood has full rank, which is the essential property we require for the CLS property. 

By considering infinitesimal geometry in $\M_n$,  we have arrived at a very similar algebraic dependence of the output of a node i.e. $\alpha_j$ to the most commonly employed affine structure in classical neural networks $h_j=\mbf{w}_j\cdot\mathbf{x}+w_{0,j}$, with the only difference being the presence of the bias $w_{0,j}$.  Such a dependence is generally required due to the elevated global importance of the origin in Euclidean spaces, while in the non-Euclidean manifold structure of $\M_n$ there is no such global notion of origin. Presence of interleaved data-independent unitaries serves a similar role in $\M_n$, which changes the orientation of the quantum feature space with respect to the observable's eigenvectors, which provide the only notion of direction in the intrinsic manifold. Thus, the bilinear aCLS-satisfying CLA map provides local selective parametric controllability, while interspersed data-independent tunable unitaries provide global flexibility, effectively acting as analogues of the bias term in classical affine maps. 

\subsection{Accessing exponentially scaling manifold dimensions} 
A major goal of quantum computing in general, and quantum machine learning in particular, is to achieve ``quantum advantage".  Under the aCLS property, a direct diagnostic of possible quantum advantage would be the scaling of possible directions of infinitesimal geometric changes and how this translates to the global scaling of such changes under the composition of such operations. Since such global effects are outside the scope of the present work, we will primarily focus on the former.  

As we have already established the geometric rigidity of data-independent unitaries and their global importance in effectively mimicking bias terms, possible geometric deformations are entirely governed by data-encoding unitary transformations regardless of the aCLS property. Evidently, restricting the generators of the encoding unitaries to the subalgebra $\oplus_{i=1}^n\mf{su}(2)$ of the product subgroup $\otimes_{i=1}^n\;SU(2)
$ caps the directions of possible deformation in $\M_n$ at $3n$. Due to the prevalence of product data-encodings, we formalise this as the following observation that captures their scaling limitations.  

\begin{observation}
If all data encoding gates in a quantum model are restricted to have generators within the  $\oplus_{i=1}^n \mf{su}(2)$ Lie algebra, the number of possible directions in $\M_n$ where one can induce selective geometric deformations is at most $3n$ even with expressive data independent unitaries with arbitrary entangling power.  
\end{observation} 

Note that the observation underscores the necessity of ``parametrised" entanglers that can continuously control the directions of some generators, thereby smoothly taking states outside the product manifold. However, for geometric deformations over all $2(2^n-1)$ real directions of the manifold $\M\cong\mathbb{C}P^{2^n-1}$, one requires an exponential number of parametrised gate operations.    

\section{Benchmarking against pure data re-upload baselines} 
\label{sec:numerical_examples}
In this section, we present numerical experiments that verify the superior performance of aCLS satisfying CLA maps compared to pure data re-uploading. 
\subsection{Datasets}  
\paragraph{Uniform hypersphere:} We first consider binary classification of two concentric five-dimensional hyperspheres embedded in $\mathbb{R}^6$.  This serves as a simple experimental verification of the underlying geometric flexibility of aCLS-satisfying CLA maps even with two qubits, where $\M_2\cong\mathbb{C}P^{3}$ has six real degrees of freedom. We construct labelled datasets $S^5_A$ and $S^5_B$ with samples containing $\mbf{x}\in\mathbb{R}^6$ and different radii normally distributed around $\bar{r}_A$ and $\bar{r}_B$ for two scenarios:
\begin{enumerate}
    \item $\bar{r}_A=0.73$ and $\bar{r}_B=0.78$ with standard deviations $\sigma_A=\sigma_B=0.05$ so that $|\bar{r}_A-\bar{r}_B|=1\sigma$
      \item $\bar{r}_A=0.72$ and $\bar{r}_B=0.78$ with $\sigma_A=\sigma_B=0.02$ so that $|\bar{r}_A-\bar{r}_B|=3\sigma$
\end{enumerate}
Using \texttt{scipy.stats.uniform\_direction.rvs} method~\cite{2020SciPy-NMeth}, we generate uniformly distributed unit vectors $\mbf{x}\in\mathbb{R}^6$ that satisfy the $S^5$ constraint $\sum_{i=1}^6 x^2_i=1$ which are then rescaled with normally distributed radii centred at $\bar{r}_A$ and $\bar{r}_B$. The total dataset for each scenario comprises 30k training, 15k validation, and 100k test samples, with equal label proportions. 

\paragraph{Binary top decay classification:} Next, we consider a binary classification task of segregating top quark decays into three final state particles via sequential two-body decays vs a direct three-body decay at the Large Hadron Collider (LHC). The sequential decay represents known physics (background) within the Standard Model of particle physics, wherein the top quark first decays to a bottom quark and a $W$ boson, with the $W$ boson then decaying to two light quarks. The three-body decay represents possible effects of yet-to-be-known physics (signal) beyond the Standard Model, where the top quark decays directly to a bottom quark and two light quarks. Since quarks are strongly interacting particles, they primarily manifest as collimated sprays of particles which are then reconstructed via specialised algorithms to form jets. 

Due to the underlying differences in the decay chain, the two classes have non-trivial topological differences\footnote{$S^2\otimes\mathbb{R}P^2$ for the sequential decay (background) and $S^5/{\sim}$ for the three-body decay (signal) with the quotient equivalence $p_{j_1}\sim p_{j_2}$ of the two jets arising from the light quark.} coupled with non-uniform probability densities via the matrix-elements. This simulated dataset was utilised in ref~\cite{Ngairangbam:2025fst} for the reconstructed level analysis of unsupervised anomaly detection.\footnote{Available at \href{https://github.com/vishalngairangbam/lat_top_ae}{https://github.com/vishalngairangbam/lat\_top\_ae}} For each event, after boosting the four-momentum $(p_x,p_y,p_x,E)$ of the final states to the rest frame of the top quark, we construct the following feature vector
\begin{equation*}
    \mathbf{x}=(p_{b}^t,\phi_b,\eta_b,m_b,p_{jj}^t,\phi_{jj},\eta_{jj},m_{jj})
\end{equation*}
where $p_{r}^t,\phi_r,\eta_r,$ and $m_r$ correspond to the transverse momentum, the azimuthal angle, the pseudorapidity,\footnote{dependent on the polar angle $\theta$ as $\eta=-\log(\tan\frac{\theta}{2})$} and mass of the object $r$, with $b$ referring to the bottom tagged jet, and $jj$ to the four-vector sum of the two jets.  The entire dataset was scaled to a range of $[-1,1]$ using $\texttt{scikit-learn's}$~\cite{sklearnapi} \textsc{MinMaxScaler}. We use 30k training, and 15k validation samples with the remaining ($\sim 575$k) used for testing the trained models. 

\paragraph{Multi-class Jet classification:}
Finally, we consider multi-class jet classification using the publicly available JetClass dataset~\cite{pmlr-v162-qu22b,qu_2022_6619768}, which presents a non-trivial challenge of identifying multi-scale structures within reconstructed jets at the LHC.  Out of the available ten classes, we chose the following five:
\begin{itemize}
    \item $q/g$: jets arising from the emission of additional particles from an energetic quark or gluon producing a single hard prong within the jet, 
    \item $t\to b\,q\,q'$: jets arising from the sequential decay of a highly energetic top quark producing a distinct pattern of three hard prongs within the jet,
    \item $H\to 4q$: jets arising from the decay of a highly energetic Higgs boson that finally results in four quarks producing four hard prongs within the jet,
    \item $Z\to qq$: jets arising from the decay products of a highly energetic $Z$ boson that decayed to two quarks, producing a two-prong jet substructure, and
    \item $H\to qq'l\nu$: jets arising from the decay of a highly boosted Higgs boson into two quarks, a lepton (an electron or a muon) and a neutrino,\footnote{The neutrinos are not directly observed but can be inferred through momentum mismatch in the final state four-vector measurement.} leading to a distinct two-prong radiation pattern with a single hard lepton within the jet.
\end{itemize}

We utilise the test dataset of each class and construct a lower-dimensional, fixed-length sample by the following preprocessing steps.
We first recluster the jet constituents into microjets of radius $R=0.2$ (the original jets were clustered with $R=0.8$) and minimum transverse momentum $p^t_m>1$ GeV via the anti-$k_t$ algorithm~\cite{Cacciari:2008gp} implemented in the $\texttt{FastJet}$~\cite{Cacciari:2011ma} package. If this reclustering produced less than five microjets, we ignore the sample and consider the first five microjets with the highest transverse momentum from those jets with more than five to construct microjet-level features $$\mbf{x}_m=(\log p^t_m,\Delta y_m,\Delta \phi_m)\quad.$$ Here, $\Delta y_m$ and $\Delta \phi_m$, respectively, are the rapidity\footnote{given in terms of four momentum components as $y=\frac{1}{2}\log\frac{E+p_z}{E-p_z}$} difference and the azimuthal angle difference of the microjet relative to the initial jet vector. In total, we extract 50k samples for each class and utilise 30k and 15k samples, with equal label proportions, for training and validation, respectively, while the remainder ($\sim 200$k samples) are used for testing the trained models. 

\subsection{Details of ansatz} 

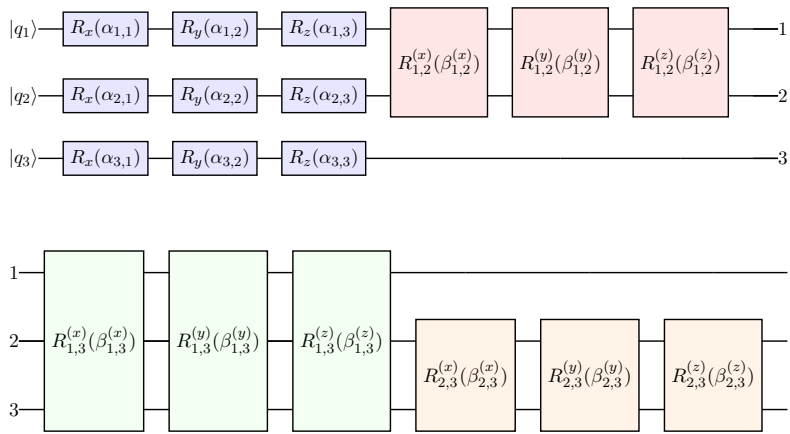
\begin{figure}[h]
    \centering
    \resizebox{0.7\textwidth}{!}{
    \begin{quantikz}
    \lvert q_1 \rangle & \gate[style={fill=blue!10}]{R_x(\alpha_{1,1})} & \gate[style={fill=blue!10}]{R_y(\alpha_{1,2})} & \gate[style={fill=blue!10}]{R_z(\alpha_{1,3})} & \gate[2, style={fill=red!10}]{R^{(x)}_{1,2}(\beta^{(x)}_{1,2})} & \gate[2, style={fill=red!10}]{R^{(y)}_{1,2}(\beta^{(y)}_{1,2})} & \gate[2, style={fill=red!10}]{R^{(z)}_{1,2}(\beta^{(z)}_{1,2})} & \qw \arrow[d, phantom, ""{coordinate, name=Z}] & \push{1} \qw \\
    \lvert q_2 \rangle & \gate[style={fill=blue!10}]{R_x(\alpha_{2,1})} & \gate[style={fill=blue!10}]{R_y(\alpha_{2,2})} & \gate[style={fill=blue!10}]{R_z(\alpha_{2,3})} & & & & \qw & \push{2} \qw \\
    \lvert q_3 \rangle & \gate[style={fill=blue!10}]{R_x(\alpha_{3,1})} & \gate[style={fill=blue!10}]{R_y(\alpha_{3,2})} & \gate[style={fill=blue!10}]{R_z(\alpha_{3,3})} & \qw & \qw & \qw & \qw & \push{3} \qw 
    \end{quantikz}}
    
    \vspace{0.8cm} 
    \resizebox{0.7\textwidth}{!}{
    \begin{quantikz}
    \push{1} \qw & \gate[3, style={fill=green!5}]{R^{(x)}_{1,3}(\beta^{(x)}_{1,3})} & \gate[3, style={fill=green!5}]{R^{(y)}_{1,3}(\beta^{(y)}_{1,3})} & \gate[3, style={fill=green!5}]{R^{(z)}_{1,3}(\beta^{(z)}_{1,3})} & \qw & \qw & \qw & \qw \\
    \push{2} \qw & \qw & \qw & \qw & \gate[2, style={fill=orange!10}]{R^{(x)}_{2,3}(\beta^{(x)}_{2,3})} & \gate[2, style={fill=orange!10}]{R^{(y)}_{2,3}(\beta^{(y)}_{2,3})} & \gate[2, style={fill=orange!10}]{R^{(z)}_{2,3}(\beta^{(z)}_{2,3})} & \qw \\
    \push{3} \qw &  &  &  &  &  & & \qw
    \end{quantikz}
    }
    \caption{The parametrised circuit block corresponding to the unitary operator $U_{SE}(\mbf{a}_{SE})$ for $n=3$ qubits.}
    \label{fig:SE_circ}
\end{figure}
As our analyses led to the analogy between data-independent unitaries and bias vectors, we utilise an encoding-and-bias-layer structure with uniform circuits for each component. Since there is no geometric gain of non-parametric gates within a fixed-ansatz structure, we utilise only parametric gates to construct two circuit blocks.
The main block utilised in all models can be written in the form
\begin{equation*}
    U_{SE}(\mbf{a}_{SE})= U_E(\mbf{a}_E)\;U_S(\mbf{a}_S)\quad, 
\end{equation*}
where $U_S(\mbf{a}_S)$ is constructed as single qubit rotations and $U_E(\mbf{a}_E)$ consists of parametrised entangling gates with $\mbf{a}_{SE}=\mbf{a}_E\oplus\mbf{a}_S$. Taking $n=3$, the circuit diagram corresponding to $U_{SE}(\mbf{a}_{SE})$ is shown in Fig.~\ref{fig:SE_circ}. In general, for any $n$ qubits, the ansatz is constructed by sequentially applying Pauli-X, Pauli-Y, and Pauli-Z rotation gates on each qubit, and then applying Ising-XX, Ising-YY, and Ising-ZZ gates in each of the two combinations according to the sequence shown in the figure.

For all experiments, we compare the performance of a Pure Data Re-uploading (PDR) model with an aCLS-satisfying model, both consisting of repeated operations of an encode-and-rotate structure. Both models utilise $U_{SE}(\mbf{a}_{SE})$ in the rotate (or bias) operations with $\mbf{a}_{SE}\equiv \w$ set as tunable parameters. In the PDR-model, we use Pauli-X rotations on each qubit, with $\din=n$ for the data-uploading ansatz. For the aCLS-model, we take $n=2$ for the hypersphere classification experiment and $n=5$ for the top and jet classification scenarios. The data encoding ansatz utilises Eq~\ref{eq:aCLS_comp} for each component of $\mbf{a}_{SE}$ to tunably distort the geometry. All considered models have three data re-uploads. 

In Appendix~\ref{sec:ansatz_unitary}, we present explicit details of the constructed unitary transformations, along with their associated weight counts and gate multiplicities. For the hypersphere classification scenario, both models contain the same number of weights in the variational ansatz, while the aCLS model uses only one quarter of the total gates required by the PDR model. In the top–quark decay classification task, the aCLS model employs approximately four times as many weights as the PDR model, but only 78\% of the PDR model's gate count. For the multi-class jet classification experiment, the aCLS ansatz has twice as many weights as the PDR model, yet requires only one quarter as many gates. While this increase in weight dimensionality for the aCLS model arises from the combined effect of its gate configuration and the linear scaling of weight multiplicity with the $\din$, across all tasks, the aCLS model has a reduced gate count compared to the PDR model. 

For all cases, we take the number of qubits $n$ to be a multiple of the output dimensions $d_{\text{out}}$ and construct the corresponding output components by partitioning the qubits into $m=k/d_{\text{out}}$ subsets say $Q_j$ for $j\in\{1,2,...,k/d\}$ as
\begin{equation}
        f_j({\bar{\w}},\x)=\sum_{i\in Q_j}\, u_i\, \quantexp{\sigma^{(i)}_z}{\psi_{\w}(\x)}\quad,
\end{equation}
with weighted tunable vector $\mbf{u}=(u_1,u_2,...,u_{n})\in\mathbb{R}^m$ adding $n$ additional weights to every model with $\bar{\w}=\w\oplus\mbf{u}$.

\begin{table}[t]
\centering 
\resizebox{0.7\textwidth}{!}{\begin{tabular}{cccccc}
\toprule
Experiment & Model
 & Wires ($n$) & AUC & Best AUC\\ 
\midrule
\multirow{2}{*}{\makecell{$S^5_A$ vs $S^5_B$ ($1\sigma$ sep.)}} &PDR  & 6 & 0.737$\pm$0.007 & 0.748\\
&aCLS  & 2 & 0.746$\pm$0.003 & 0.748 \\
\midrule 
\multirow{2}{*}{\makecell{$S^5_A$ vs $S^5_B$ ($3\sigma$ sep.)}}  &PDR  & 6 & 0.951$\pm$0.013 & 0.964\\
&aCLS  & 2 & 0.965$\pm$0.005 & 0.973 \\
\midrule\multirow{2}{*}{Top Decay}&PDR  & 8 & 0.751$\pm$0.001 & 0.752 \\
&aCLS  & 5 & 0.795$\pm$0.003 & 0.798 \\
\bottomrule
\end{tabular}
}
\caption{The averaged AUC and its standard deviation across five runs, and the best AUC for the binary classification scenarios. The PDR(Pure data re-uploads)-model corresponds to $\text{rank}(\mbf{J}_\mc{W}(\x))=0$ data encoding circuit blocks, while the aCLS-model corresponds to those with almost complete data encodings based on Eq~\ref{eq:aCLS_comp}. We see that even with six real degrees of freedom in $\M_2\cong \mathbb{C}P^{3}$ that matches the minimal embedding dimensions of $S^5$, the aCLS-model consistently outperforms the kernel-like PDR model.}
\label{tab:binary_auc} 
\end{table}

All models are implemented in \texttt{PennyLane}~\cite{Bergholm:2018cyq} with the \texttt{PyTorch}~\cite{10.5555/3454287.3455008} backend. 
To serve as a uniform benchmark for the geometric efficiency of the models, the initial reference state $\ket{\psi_0}$ is constructed with randomly sampled real and distinct amplitudes (via the \textsc{AmplitudeEmbedding} class), which is kept the same for all experiments by fixing the seed. We construct one-dimensional outputs for binary classification tasks and use binary cross-entropy loss, whereas for the multi-class classification task, we use categorical cross-entropy with a five-dimensional output vector. The models are trained with the \texttt{Adam}~\cite{DBLP:journals/corr/KingmaB14} optimiser with an initial learning rate of 0.001, which decays by a factor of 0.1 if the validation loss has not decreased for three epochs. We train all models five times from random weight initialisation for up to 300 epochs, with a batch size of 128 and an early stopping criterion triggered if the validation loss has not decreased for 20 epochs. The results are inferred on the test dataset using weights saved from the epoch with the lowest validation loss.

\begin{figure}[t!]
\centering 
\includegraphics[width=0.49\linewidth]{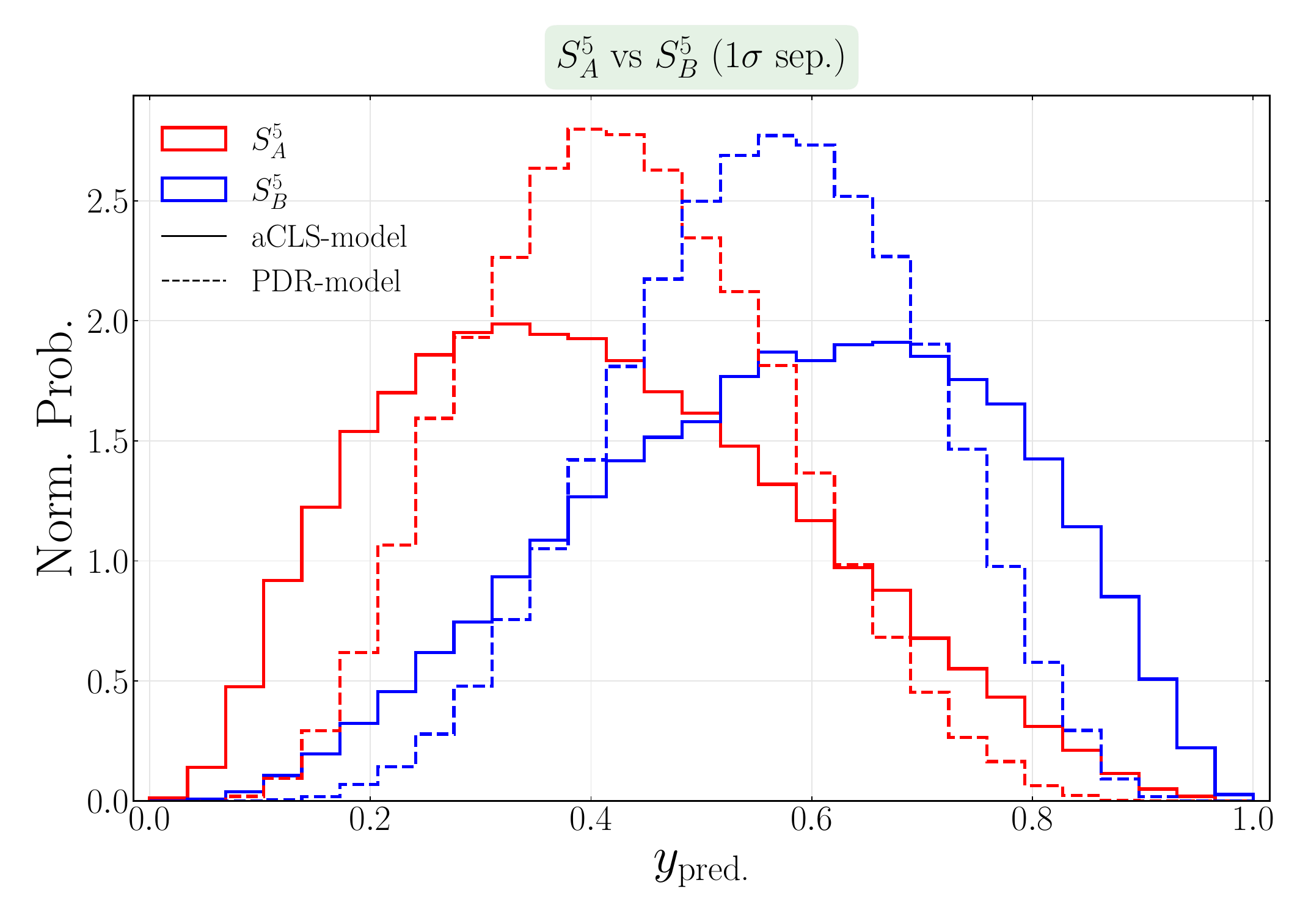}
    \includegraphics[width=0.49\linewidth]{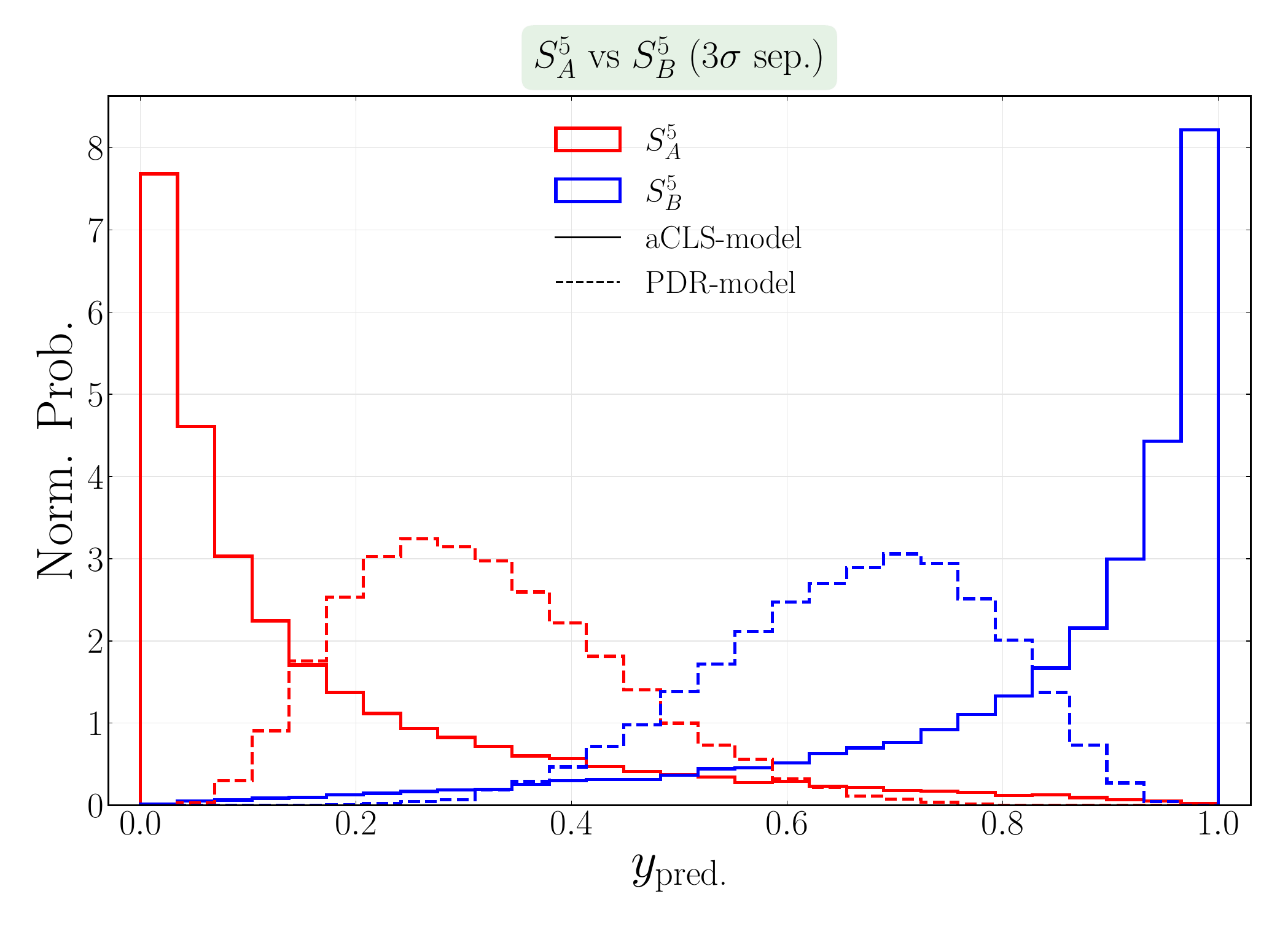}  
    \includegraphics[width=0.49\linewidth]{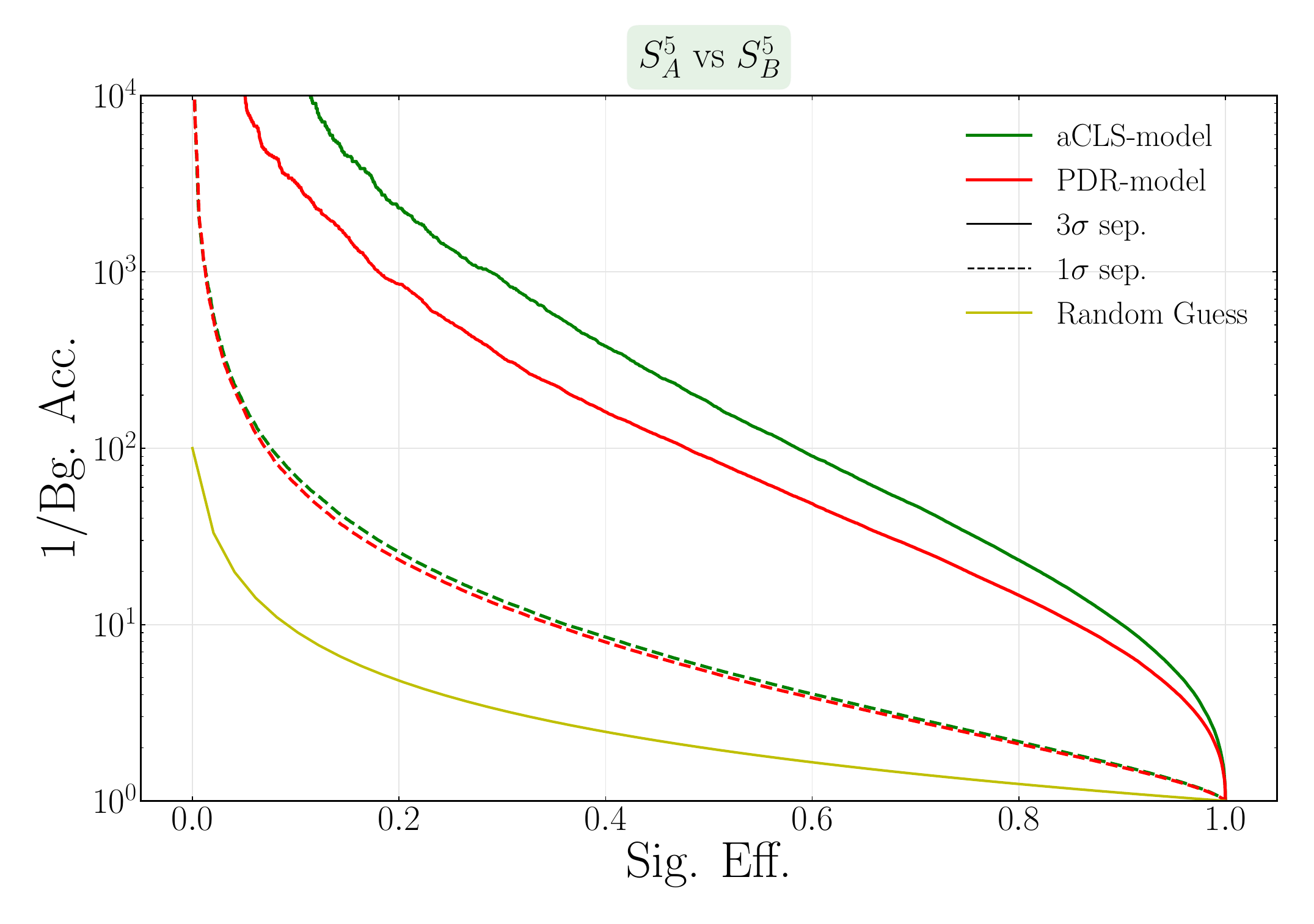}
\caption{Output histograms for the $S^5_A$ vs $S^5_B$ binary classification experiment comparing the best performing aCLS and PDR models for the $|\bar{r}_A-\bar{r}_B|=3\sigma$ case (top-right) and $|\bar{r}_A-\bar{r}_B|=1\sigma$  (top-left), along with their corresponding ROC curves (bottom).}
\label{fig:hypersphere} 
\end{figure}
\begin{figure}[t]
\centering 
        \includegraphics[width=0.49\linewidth]{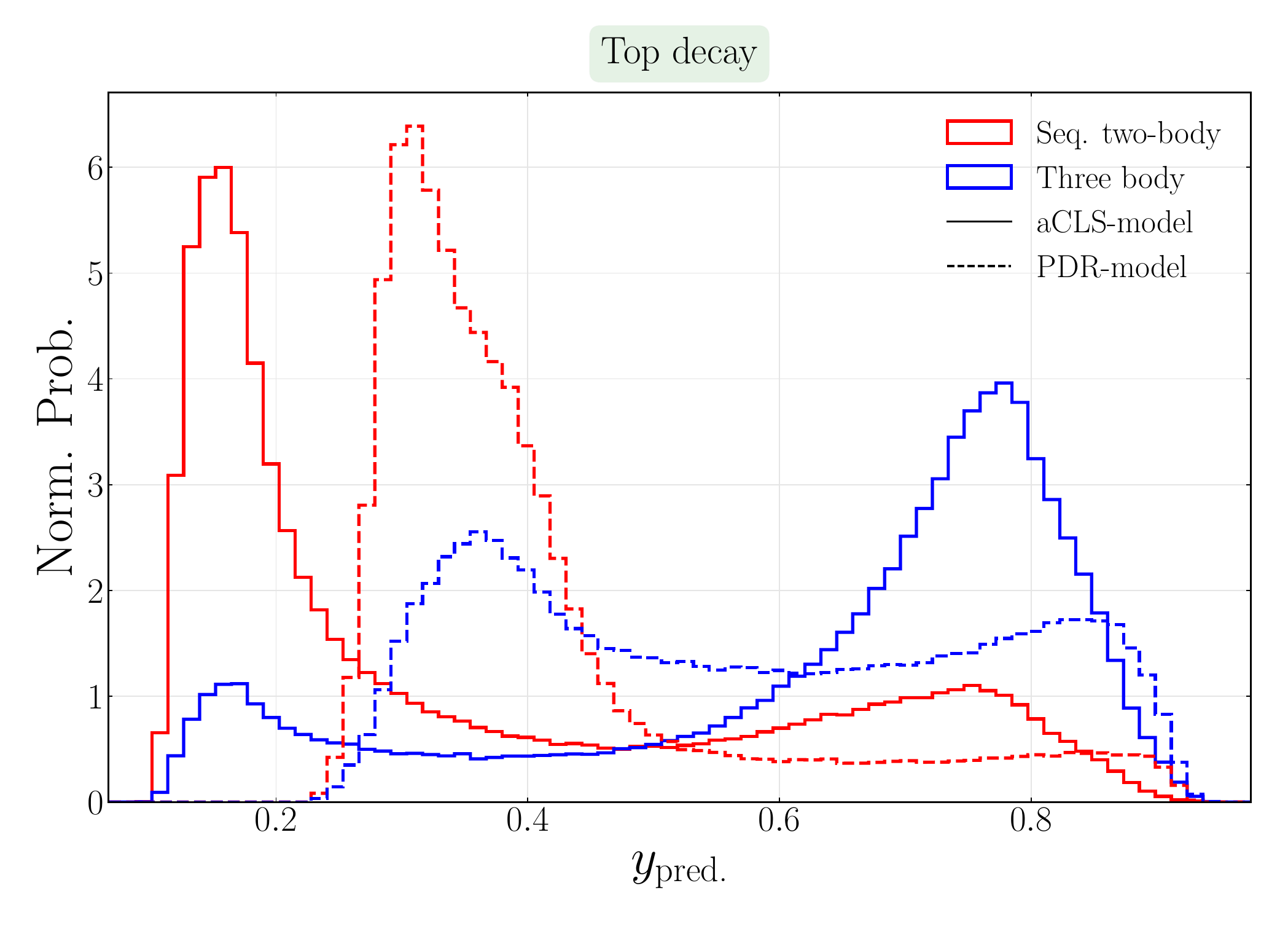}
    \includegraphics[width=0.49\linewidth]{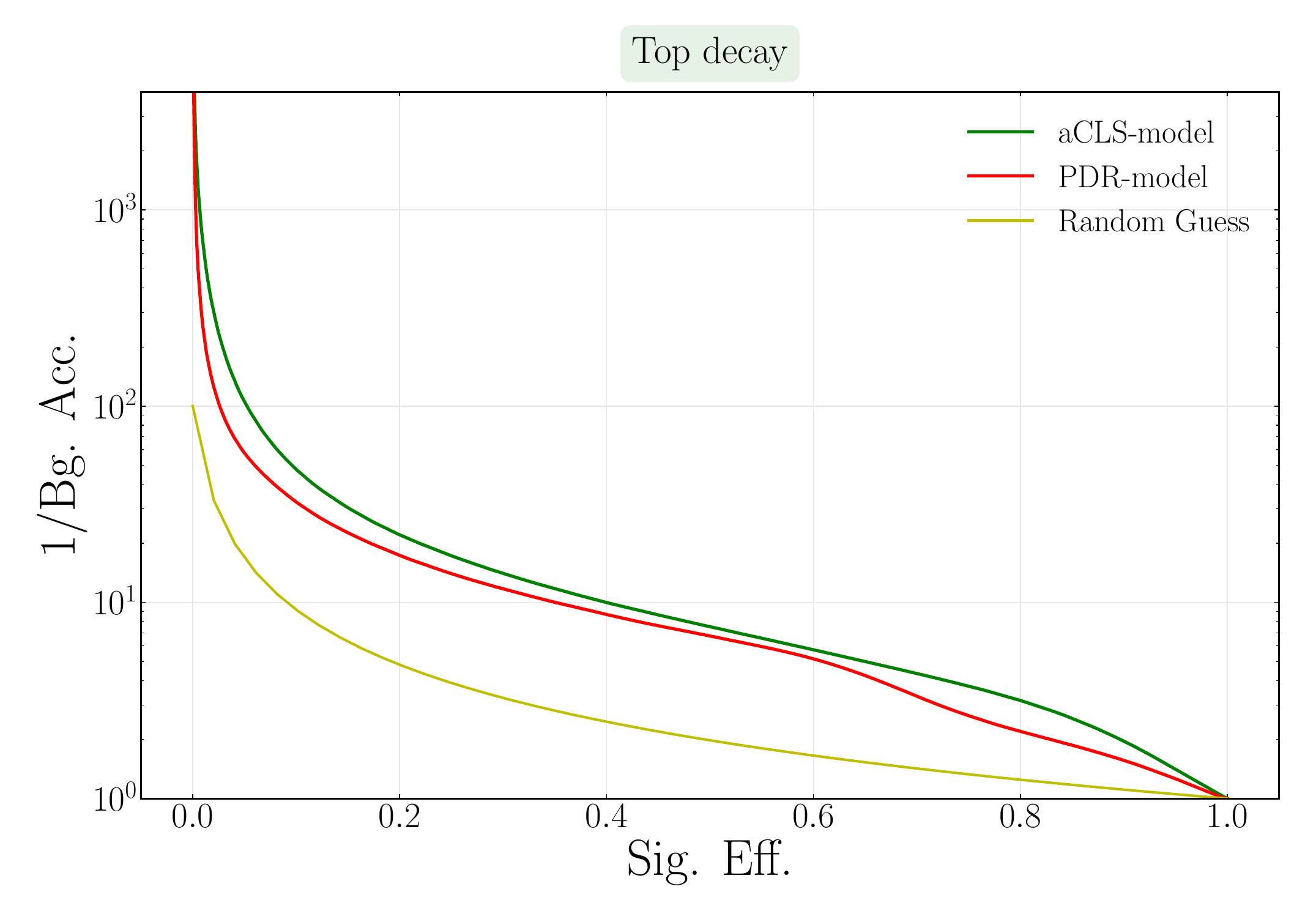}  
    \caption{Output histogram (left) of the best-performing aCLS and PDR models for the top-decay classification and the corresponding ROC curves (right).}
\label{fig:top_dec} 
\end{figure}

\subsection{Performance comparison}
In Tab.~\ref{tab:binary_auc}, we show the Area-Under-the-Curve (AUC) of the Receiver Operator Characteristics (ROC) averaged over five runs and its standard deviation for the binary classification scenarios along with AUCs of the best run. For this best-performing model, the output histograms and the ROC curve are shown in Figs.~\ref{fig:hypersphere} and~\ref{fig:top_dec}. For the hypersphere classification scenarios, we see that even with two qubits, the aCLS-model outperforms the six-qubit PDR-model in both the low-noise and high-noise settings. Compared to the toy datasets, the difference between the two models becomes more pronounced for the top decay classification data which in addition to its non-trivial topological difference also has non-uniform distributions. 

\begin{figure}[t!]
\centering 
\includegraphics[width=0.45\textwidth]{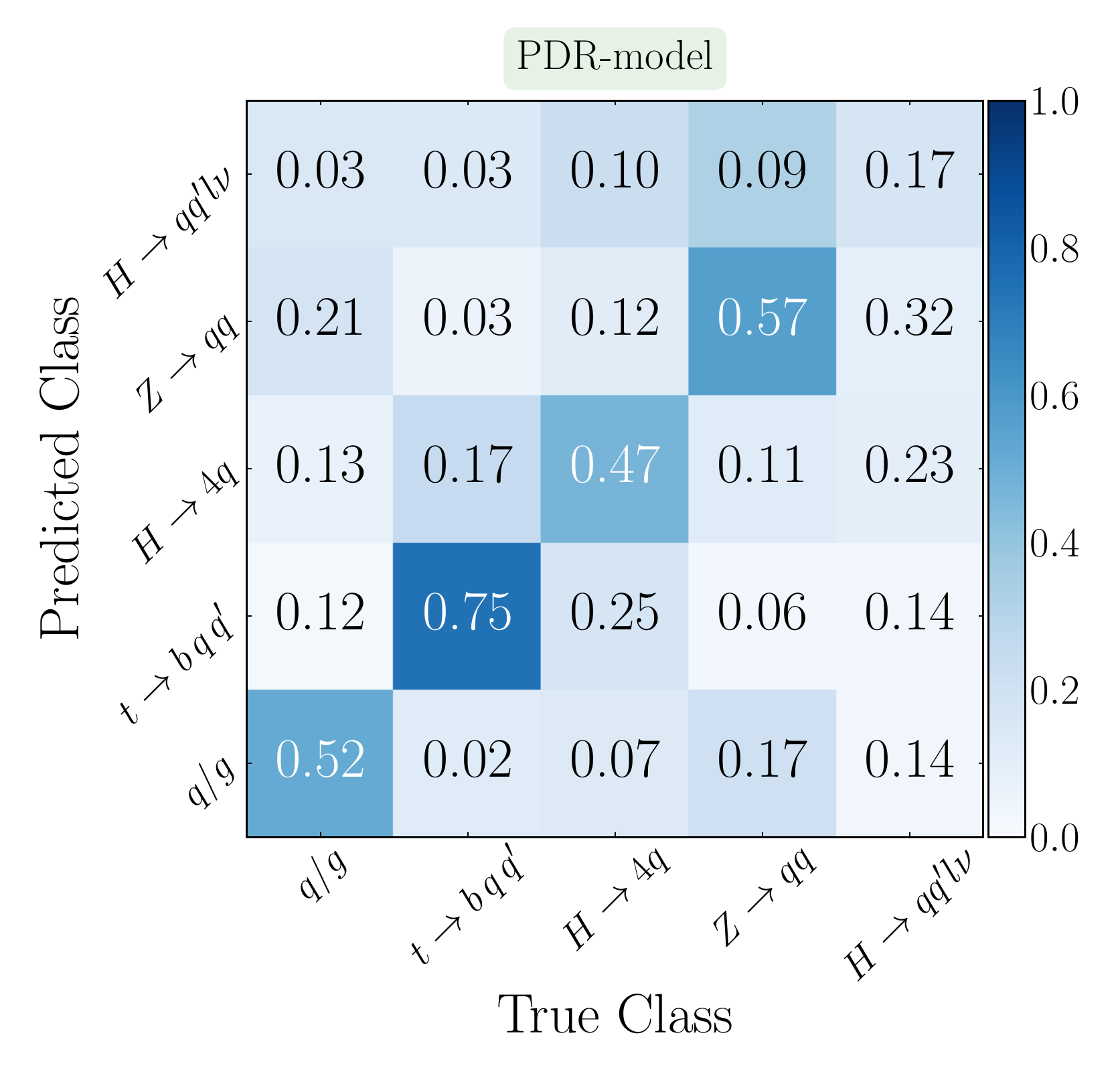}
\includegraphics[width=0.45\textwidth]{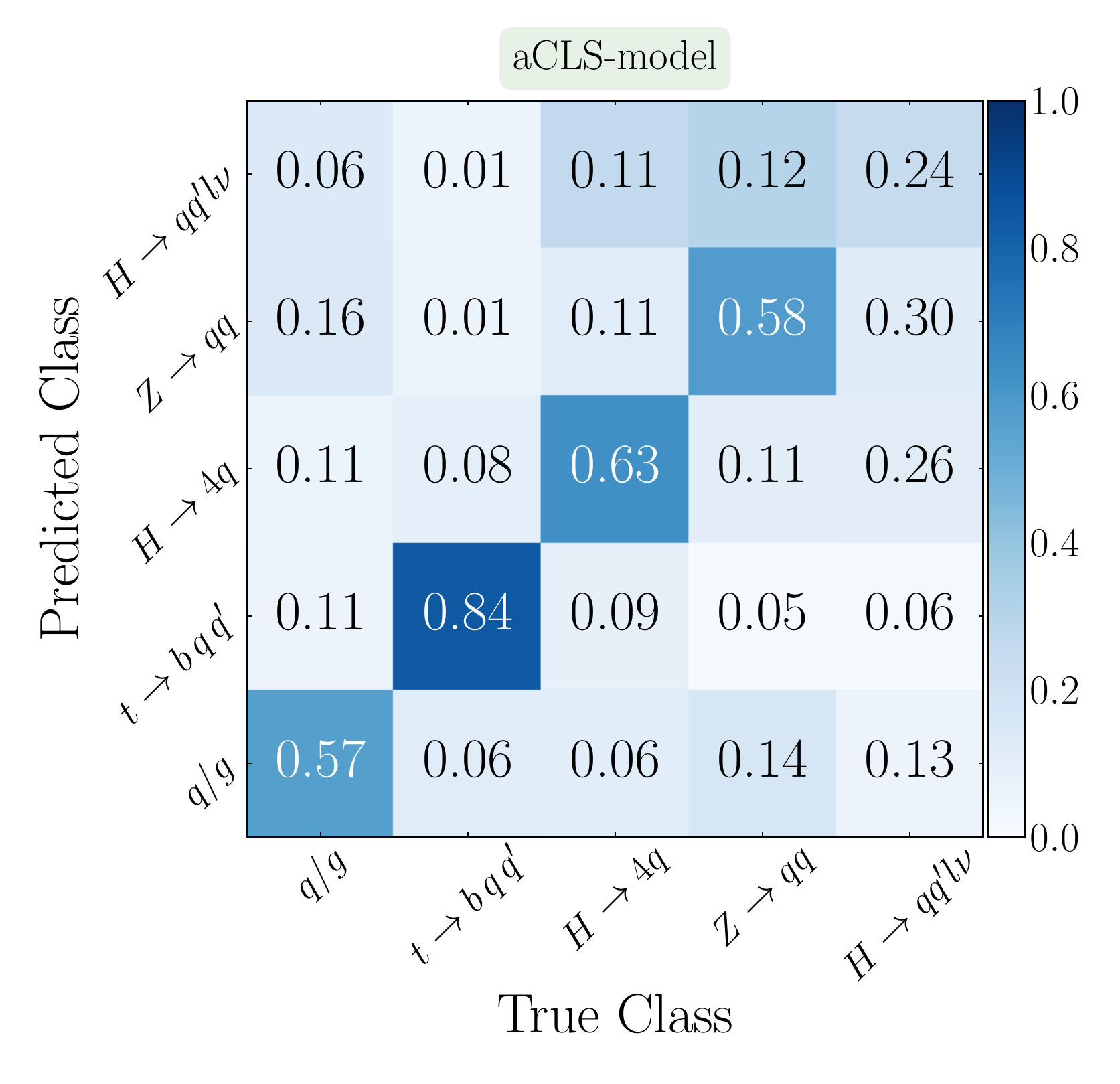}
\caption{Averaged confusion matrices normalised by true labels of the aCLS (right) and PDR (left) models for the multi-class jet classification scenario. }
\label{fig:jetclass} 
\end{figure}

In Fig.~\ref{fig:jetclass}, we show the averaged (over different training runs) confusion matrix of each model normalised over true labels for the multi-class jet classification experiment. Across all classes, the aCLS-model consistently achieves higher accuracy (diagonal elements) than the PDR-model. Comparing across columns, we see that the aCLS-model consistently has lower misclassification proportions across all true classes, whereas the PDR-model has lower proportions in at most one class. Among the five classes, this occurs three times for $H\to qq'l\nu$ jets, which nevertheless remains the class with the poorest overall accuracy. To quantify this numerically, we show the one-vs-all averaged AUC score (along with the maximum) of the ROC curve in Tab.~\ref{tab:multi-class_jetclass} where the aCLS-model consistently outperforms the PDR model.

\begin{table}[t]
\centering 
\resizebox{0.6\textwidth}{!}{\begin{tabular}{ccccc}
\toprule
Class label  & Model & Wires ($n$)  & AUC & Best AUC \\
\midrule
\multirow{2}{*}{$q/g$} &PDR & 15 & 0.813$\pm$0.003 & 0.816 \\
&aCLS & 5 & 0.834$\pm$0.002 & 0.836 \\

\midrule
\multirow{2}{*}{$t\to b\,q\,q'$}& PDR & 15 & 0.893$\pm$0.003 & 0.896 \\
&aCLS & 5 & 0.951$\pm$0.001 & 0.951 \\

\midrule
\multirow{2}{*}{$H\to 4q$}&PDR & 15 & 0.777$\pm$0.005 & 0.782 \\
&aCLS & 5 & 0.848$\pm$0.001 & 0.850 \\

\midrule
\multirow{2}{*}{$Z\to qq$}&PDR & 15 & 0.797$\pm$0.001 & 0.797 \\
&aCLS & 5 & 0.815$\pm$0.001 & 0.816 \\

\midrule 
\multirow{2}{*}{$H\to qq'l\nu$}&PDR & 15 & 0.676$\pm$0.004 & 0.681 \\
&aCLS & 5 & 0.724$\pm$0.002 & 0.726 \\
\bottomrule
\end{tabular}}
\caption{The averaged one-vs-all AUC and its standard deviation across five runs, and the best AUC for each class of the multi-class jet classification experiment.  As explained in the text, the total gates utilised in the aCLS-model is a quarter of those utilised in the PDR-model.}
\label{tab:multi-class_jetclass} 
\end{table}

\section{Summary and Conclusions}
\label{sec:summary_conclusions}

In this work, we asked what it really means for a quantum neural network layer to learn features, rather than merely apply a trainable but essentially rigid transformation to an already encoded dataset. We took a geometric viewpoint and regarded the classical input space as embedded into the pure-state manifold $\M_n\simeq \mathbb{C}P^{2^n-1}$ through an encoding map, and we evaluated an ansatz by whether training can \emph{adaptively deform} the local geometry of the embedded data manifold. This shifts the focus from the reachability of states to the learnability enabled by controllable, data-dependent geometric actions.  

To make this precise, we connected circuit design to infinitesimal unitary motion on $\M_n$ via Lie-algebra directions. We introduced Classical-to-Lie-algebra maps that associate the input $\x$ and the trainable parameters $\w$ to an effective generator
\[
\Gamma(\w,\x)=\sum_{j=1}^k \alpha_j(\w_j,\x)\,G_j,
\]
with generators $G_j\in\mathfrak{su}(2^n)$. This construction allowed us to express architectural capability as an accessible set of tangent directions and their trainable modulation over the data manifold.

On this basis, we defined almost Complete Local Selectivity as an almost-everywhere rank condition expressing two requirements at once: we demand completeness in weight space so that parameters can independently control the available directions, and we demand local selectivity in data space so that the action varies across inputs in a way that training can tune. In practice, this pushes us toward nontrivial joint dependence on $(w,x)$, for example, through bilinear forms $\alpha_j(\w_j,\x)=\w_j\cdot \x$, instead of separating data and weights into disjoint circuit blocks.

With this criterion in hand, we diagnose why two common constructions fall short. If we append a stack of trainable unitaries that does not depend on the input, then we can be expressive in the sense of exploring many unitaries, yet we still act as an isometry on the encoded manifold. Such layers preserve unitarily invariant distances and therefore behave like learnable rigid rotations rather than learnable deformations.

At the other extreme, if we rely on pure data re-uploading without tunable data dependence, then we obtain selectivity without genuine tunability. The deformation pattern is fixed by the encoding itself, and increasing depth does not automatically enable adaptive feature learning. We also highlight associated structural restrictions that appear naturally in this setting.

We then translated these geometric insights into concrete design rules for ansatzes. Most importantly, to exploit the large real dimension of $\M_n$, we must include parametrised entangling directions. Fixed entanglers alone do not provide continuously controllable geometric directions, and restricting ourselves to purely local $\oplus_i\mathfrak{su}(2)$ actions caps the available deformations at a scale set by $n$ rather than by $2^n$.

An aCLS-satisfying re-uploading construction implements these requirements by coupling trainable weights to data-dependent generators, while retaining a gate structure that remains feasible at small qubit counts.

We empirically validated the framework using supervised learning benchmarks that compare an aCLS-satisfying ansatz to a pure data-reuploading baseline. On a controlled-geometry task based on concentric hyperspheres, we observe clear gains in AUC, including regimes in which we achieve better performance with fewer qubits than the baseline. On collider-motivated datasets, we see the same pattern. For a binary top-decay task, the aCLS ansatz yields a noticeable AUC improvement over the pure re-uploading reference, and on the multi-class JetClass benchmark, we observe consistent one-vs-all AUC improvements across all classes, in line with the improved confusion-matrix structure.

At the same time, we found that the geometric design principles often translate into better circuit efficiency. In representative setups, we substantially reduced the number of parametrised gates, often to around a quarter of the baseline, while maintaining competitive parameter counts and improving predictive performance.

Overall, we provided an operational criterion for when a QNN layer can perform feature learning: it must provide a sufficiently rich set of parametrically controllable Lie-algebra directions, and it must couple these directions to the data in a trainable way so that optimisation can reshape the local geometry of the embedded manifold. The CLA-map formalism and the aCLS criterion provide a compact checklist for this goal and, in geometric terms, explain why commonly used architectures can plateau as learnable isometries or as fixed deformations. Our results show that enforcing controllable geometry improves both performance and resource efficiency on the benchmarks we study, and motivate extending this geometric learning viewpoint beyond the present pure-state closed-system setting.
\section*{Acknowledgements}
This work was funded by the STFC under grant ST/X003167/1.
\appendix 
\section{Proofs and worked-out examples}
\label{app:proofs} 

\subsection{Proposition~\ref{prop:gate_tun_sel_act}}
\label{sec:gate_tun_sel_act} 
Since the Fubini-Study distance is a monotonic implicit function of fidelity, its analytic dependence on weights and input samples is determined by fidelity through application of the chain rule.
Define $\Delta \alpha=\alpha(\w,\x_1)-\alpha(\w,\x_2)$ and write the fidelity as
\begin{equation*}
\begin{split}
    F(\ket{\psi_\w(\x_1)},\ket{\psi_\w(\x_2)})&=|\bra{\psi(\x_1)}e^{i\;\Delta \alpha\;G}\ket{\psi(\x_2)}|^2 \\
    &=\bra{\psi(\x_1)}e^{i\;\Delta \alpha\;G}\ket{\psi(\x_2)}\;\bra{\psi(\x_2)}e^{-i\;\Delta \alpha\;G}\ket{\psi(\x_1)}
    \end{split} 
\end{equation*}
Using the completeness relation of the eigenvectors of the Hermitian generator $G$, say $\sum_a\;\ket{\lambda_a}\bra{\lambda_a}=\hat{I}$, we have 
\begin{equation*}
    \begin{split}
        \bra{\psi(\x_1)}e^{i\;\Delta \alpha\;G}\ket{\psi(\x_2)}=&\sum_{a}\bra{\psi(\x_1)}e^{i\;\Delta \alpha\;G}\ket{\lambda_a}\braket{\lambda_a}{\psi(\x_2)}\\
        =&\sum_a\;e^{i\;\Delta\alpha \;\lambda_a} \braket{\psi(\x_1)}{\lambda_a}\braket{\lambda_a}{\psi(\x_2)}\\
        =&\sum_a\;e^{i\;\Delta\alpha \;\lambda_a}\; c^*_a(\x_1)\;c_a(\x_2)
    \end{split}
\end{equation*}
where we have defined $c_a(\x_{i})=\braket{\lambda_a}{\psi(\x_i)}$. Thus, we get 
\begin{equation*}
    |\bra{\psi(\x_1)}e^{i\;\Delta \alpha\;G}\ket{\psi(\x_2)}|^2=\sum_{a,b} e^{i\;\Delta\alpha\;(\lambda_a-\lambda_b)}\; c^*_a(\x_1)\;c_a(\x_2)\;c_b(\x_1)\;c^*_b(\x_2)\quad.
\end{equation*}
Differentiating with $w_j$ we get
\begin{equation}
\label{eq:fid_diff_weight} 
\begin{split} 
    \frac{\partial F}{\partial w_j}=&i\sum_{a,b}\frac{\partial\Delta \alpha}{\partial w_j}(\lambda_a-\lambda_b)\;e^{i\Delta\alpha\;(\lambda_a-\lambda_b)}\;c^*_a(\x_1)\;c_a(\x_2)\;c_b(\x_1)\;c^*_b(\x_2)\quad.
    \end{split}
\end{equation}
Hence as long as the two states $\ket{\psi(\x_1)}$ and $\ket{\psi(\x_2)}$ are not orthogonal to at least two distinct eigenvectors of $G$,  $\frac{\partial \Delta\alpha}{\partial w_j}$ decides the dependence of $D_{FS}$. From the Mean Value theorem, due to the assumption of smoothness of  $\alpha(\w,\x)$, we have some $x_{l,0}\in[x_{l,1},x_{l,2}]$, such that 
\begin{equation*}
    \left.\frac{\partial^2\alpha}{\partial x_l\;\partial w_j}\right|_{x_{l,0}}=\frac{1}{x_{l,2}-x_{l,1}}\frac{\partial \Delta \alpha}{\partial w_j}
\end{equation*}
This completes the proof.

\subsection{Fidelity of selective tunable gate on 1D domain}
\label{sec:fid_1d_domain}
We examine a simple one-dimensional case to elucidate the geometric structure characterising selectively and non-selectively tunable gates. Suppose we have a set of Pauli-X-rotated encoded states   
\begin{equation}
    \ket{\psi(x)}\equiv e^{\frac{i}{2}\,x\,\sigma_x}\ket{0}=\cos\left(\frac{x}{2}\right)\ket{0}-i\sin\left(\frac{x}{2}\right)\ket{1} 
\end{equation}
with $\X$ a subinterval of $(-\pi,\pi]$. Under a Pauli-Z rotation gate $U(\alpha)=e^{\frac{i}{2}\,\alpha\,\sigma_z}$, consider $\alpha(w,x)=w$, which acts as a non-selective and tunable transformation  on the states as 
\begin{equation*}
    \ket{\psi_w(x)}=e^{\frac{i}{2}\,w\,\sigma_z}\ket{\psi(x)}\quad.
\end{equation*}
The inner product between any two states $\ket{\psi_w(x_1)}$ and $\ket{\psi_w(x_2)}$ evaluates to $$\braket{\psi_w(x_1)}{\psi_w(x_2)}=\bra{\psi(x_1)}e^{-i\,w\,\frac{\sigma_z}{2}}e^{i\,w\,\frac{\sigma_z}{2}}\ket{\psi(x_2)}=\braket{\psi(x_1)}{\psi(x_2)}\quad,$$ making the fidelity (and hence the Fubini-Study metric) independent of $w$. 

Taking $\alpha(w,x)=wx$ minimally satisfies $\frac{\partial^2\alpha}{\partial x\partial w}=1$ making it selective and tunable.  The evolved state under $e^{\frac{i}{2}\,wx\,\sigma_z}$ evaluates to 
\begin{equation*}
    \ket{\psi_w(x)}=e^{-i\,wx/2}\cos\left(\frac{x}{2}\right)\ket{0}-i\;e^{i\,wx/2}\;\sin\left(\frac{x}{2}\right)\ket{1}\quad, 
\end{equation*}
from which one can work out the inner product between two states $\ket{\psi_w(x_1)}$ and $\ket{\psi_w(x_2)}$ as 
\begin{equation*}
    \begin{split}
      \braket{\psi_w(x_1)}{\psi_w(x_2)}&=  \cos\left(\frac{x_1}{2}\right)\cos\left(\frac{x_2}{2}\right) e^{i\,w(x_1-x_2)/2}+\sin\left(\frac{x_1}{2}\right)\sin\left(\frac{x_2}{2}\right)e^{-i\,w(x_1-x_2)/2}\\
      &
    \end{split}
\end{equation*}
Expanding the exponentials and simplifying the trigonometric relations, we get the fidelity as
\begin{equation*}
\begin{split} 
    F(\ket{\psi_w(x_1)},\ket{\psi_w(x_2)})=\cos^2&\left(\frac{w}{2}(x_1-x_2)\right)\cos^2\left(\frac{1}{2}(x_1-x_2)\right)\\&+\sin^2\left(\frac{w}{2}(x_1-x_2)\right)\cos^2\left(\frac{1}{2}(x_1+x_2)\right)
\end{split} 
\end{equation*}

Evaluating the partial derivative with respect to $w$ and simplifying the expression with trigonometric identities, we get 
\begin{equation*}
    \frac{\partial F}{\partial w}=-\frac{x_1-x_2}{2}\sin\left(w(x_1-x_2)\right)\;\sin(x_1)\sin(x_2)\quad.
\end{equation*}
We see that the right-hand side evaluates to zero whenever $x_1$ or $x_2$ goes to zero due to their corresponding sine factors. Going back to Proposition~\ref{prop:gate_tun_sel_act}, this arises because $x=0$ does not rotate the initial state $\ket{0}$, which, being an eigenvector of the Hermitian generator $\sigma_z$, is naturally orthogonal to its only other distinct eigenvector $\ket{1}$.  Such sets are lower-dimensional subsets (here, isolated points in the 1D domain $\X$) of the domain and therefore occur in a negligible subset.

\subsection{Rigidity of data-independent unitaries} 
\label{sec:rigid_isom}
Consider that we have a two-dimensional domain $\X$ that is a subset of $(-\pi,\pi]^2$ which undergoes an angle encoding via Pauli-X rotation gates on two qubits as 
\begin{equation}
\label{eq:input_enc} 
    \ket{\psi(\x)}=e^{\frac{i}{2}\,x_1\,\sigma_x}\ket{0}\otimes e^{\frac{i}{2}\,x_2\,\sigma_x}\ket{0}\quad.
\end{equation}
 Since generic applications of ML on 2D domain require non-trivial interactions between the input variables $x_1$ and $x_2$, consider that we have an ansatz that consists of parametrically controllable interaction based on different Ising interactions, say $\mbf{K}=\{\sigma_x\otimes \sigma_x,\sigma_y\otimes\sigma_y,\sigma_z\otimes\sigma_z\}$ as 
\begin{equation}
\label{eq:tun_unit}
S(\w)=e^{\frac{i}{2}\,w_1\,\sigma_x\otimes \sigma_x}\;e^{\frac{i}{2}\,w_2\,\sigma_y\otimes \sigma_y}\;e^{\frac{i}{2}\,w_3\,\sigma_z\otimes \sigma_z}\quad,
\end{equation} 
with $\w=(w_1,w_2,w_3)$. This has the corresponding CLA map
\begin{equation*}
    \Gamma(\w)=w_1\,\sigma_x\otimes \sigma_x+ w_2\,\sigma_y\otimes \sigma_y+w_3\;\sigma_z\otimes \sigma_z
\end{equation*}
with $\mbf{a}(\w)=\w$, making the weight Jacobian $\mbf{J}_\mc{W}=I_3$ which has rank three.

Additionally, if there is a non-parametric gate in the ansatz, like CNOT, we have the unitary circuit 
$$S'(\w)=S(\w)\,C_X\quad,$$
 $C_X=e^{-\frac{i}{2}\;\frac{\pi}{2} \gamma}$ being the CNOT gate's unitary transformation with the generator $\gamma$, given as $$\gamma=I_2\otimes I_2-I_2\otimes\sigma_x-\sigma_z\otimes I_2+\sigma_z\otimes\sigma_x\quad$$
in the Pauli string basis. Thus, the linear term in the expansion of the unitary transformation $S'(\w)$ has a $\gamma$ dependent term 
\begin{equation*}
\Lambda(\w)=  \frac{i}{2}\left(\Gamma(\w)-\frac{\pi}{2}\gamma\right)\quad,
\end{equation*}
which, however, does not add to the controllable geometric directions under the CLA map.  Note that globally, the accessible range of unitary transformations under $S(\w)$ is in a one-to-one correspondence for different values of $\w$ with $S'(\w)$ elements obtained by fixed right-translations with $C_X$.

If, on the other hand, we had two repeated applications of the same unitary
\begin{equation*}
    U(\w,\w')=S(\w)\,S(\w')\quad,
\end{equation*}
with new independent weights $\w'=(w'_1,w'_2,w'_3)$, 
the underlying vector subspace of the Lie algebra is  still described by the generator set $\mbf{K}$. This can be seen by from the CLA map
\begin{equation*}
      \Gamma'(\w,\w')=(w_1+w'_1)\,\sigma_x\otimes \sigma_x+ (w_2+w'_2)\,\sigma_y\otimes \sigma_y+(w_3+w'_3)\;\sigma_z\otimes \sigma_z\quad,
\end{equation*}
where the image is isomorphic to $\GL{3}{\mathbb{R}}$, thereby making three real degrees of freedom locally redundant.

\section{Unitary transformation of ansatz and gate counts}
\label{sec:ansatz_unitary}
For any $n$, the single qubit rotation consists of sequential rotations of Pauli-X, Pauli-Y and Pauli-Z rotations in each qubit 
\begin{equation*}
U_S(\mbf{a}_S)=\bigotimes_{i=1}^n\left[\;e^{\frac{i}{2}\alpha_{i,3}\;\sigma_z}\;e^{\frac{i}{2}\alpha_{i,2}\;\sigma_y}\;e^{\frac{i}{2}\alpha_{i,1}\;\sigma_x}\right]\quad,
\end{equation*}
with the vector map $\mbf{a}_S=(\alpha_{1,1},\alpha_{1,2},\alpha_{1,3},...,\alpha_{n,1},\alpha_{n,2},\alpha_{n,3})\in\mathbb{R}^{3n}$.  

$U_E(\mbf{a}_E)$ is built out of Ising-XX, Ising-YY, and Ising-ZZ parametrised entanglers $R^{(a)}_{i,j}(\alpha)=e^{\frac{i}{2} \alpha\;G^{(a)}_{i,j}}$, $a\in\{x,y,z\}$, that acts on the $i^{th}$ and $j^{th}$ qubits. For some $n$ qubit systems, these gates are organised to give the following unitary transformation
\begin{equation*}
    U^\dagger_E(\mbf{a}_E)=\prod_{i=1}^{n-1}\prod_{j=i+1}^n\; R^{(x)}_{i,j}(-\beta^{(x)}_{i,j})\;R^{(y)}_{i,j}(-\beta^{(y)}_{i,j})\;R^{(z)}_{i,j}(-\beta^{(z)}_{i,j})\quad,
\end{equation*}
where $\mbf{a}_E=(\beta^{(x)}_{1,2},\beta^{(y)}_{1,2},\beta^{(z)}_{1,2},\beta^{(x)}_{1,3},\beta^{(y)}_{1,3},\beta^{(z)}_{1,3},...,\beta^{(x)}_{n-1,n},\beta^{(y)}_{n-1,n},\beta^{(z)}_{n_-1,n})\in\mathbb{R}^{3n_c}$ and $n_c=\binom{n}{2}$. Thus, the $\mbf{G}$-subspace dimensions is $k\equiv k_{SE}(n)=3(n+\binom{n}{2})$.  

Beyond the geometric consequences of pure data-encodings, there is no principled way to encode classical data (with no implicit bias) via non-commuting gate operations in the absence of any tunable parameter dependence. Therefore, we use single Pauli-X rotations on each qubit for the pure data re-upload model, whose unitary transformation can be written as 
$$U_X(\mbf{a}_X)=\bigotimes_{i=1}^n\;e^{\frac{i}{2}\alpha_i\;\sigma_x}\quad,$$
where $\mbf{a}_X=(\alpha_1,\alpha_2,...,\alpha_n)\in\mathbb{R}^n$ with $k\equiv k_X(n)=n$.

For the PDR model (i.e. $\DJac{\w}=0$), we use $n=\din$, with each layer implemented as  the unitary transformation 
\begin{equation*}
    F_l(\mbf{v}^{(l)},\x)=U_{SE}(\mbf{v}^{(l)})\;U_X(\mbf{x})\;\quad.
\end{equation*}
Here, $\mbf{a}_{SE}\equiv\mbf{v}^{(l)}\in\mathbb{R}^{k_{ES}}$ are tunable, and $\mbf{a}_X\equiv\mbf{x}$ are fixed one-to-one maps to the principal domain of the Pauli-X rotation gates. The total number of gates for each $l$ evaluates to 
\begin{equation*}
    G_p(n)=k_{SE}(n)+k_X(n)\quad,
\end{equation*}
while the total weights per $l$ evaluates to
\begin{equation*}
    d_{w,p}(n)=k_{SE}(n)\quad. 
\end{equation*}
With similar data-independent unitary $U_{SE}(\mbf{v}^{(l)})$, the aCLS-satisfying model's layers have unitary transformations of the form
\begin{equation}
    T_l(\mbf{v}^{(l)},\w^{(l)},\x)=U_{SE}(\mbf{v}^{(l)})\;U_{SE}(\w^{(l)},\mbf{x})\quad,
\end{equation}
where each component of $\mbf{a}_{SE}$ has the structure $\alpha^{(l)}_i=\mbf{w}^{(l)}_j\cdot \mbf{x}$. Therefore, per $l$, the model consists of \begin{equation*}
G_t(n)=2 k_{SE}(n) \quad
\end{equation*}
gates, while the weight space has dimensions
\begin{equation*}
    d_{w,t}(n,\din)=(\din+1)\;k_{SE}(n)\quad.
\end{equation*}

For the $S^5_A$ vs $S^5_B$ classification with $\din=6$, we take $n=2$ for the aCLS-model, with $n=6$ for the PDR model so that $d_{w,t}(2,6)=d_{w,p}(6)=63$, serving as a direct comparison of the tunable parameter efficiency of the models. Moreover, note that the ratio of gate multiplicity is $G_t(2)/G_p(6)=18/69\approx26\%$. Thus, while the weights are identical when we use the same $L$ for both models, the considered aCLS model utilises approximately a quarter of the parametrised gate operations of the PDR model. 

For the binary top decay classification, we take $n=5$ for the aCLS model, while $n=\din=8$ for the PDR model with the weight ratio $d_{w,t}(5,8)/d_{w,p}(8)=3.75$ and the gate ratio $G_t(5)/G_p(8)\approx 78\%$.  As the aCLS model's weights scale linearly with $\din$, increasing input dimensions leads to a faster increase in the number of weights compared to the PDR model. However, the aCLS model still has fewer gates per layer than the PDR model. Similarly, for multi-class jet classification, we consider an $n=5$ aCLS model, with the PDR model having $n=\din=15$ qubits. This leads to the weight ratio $d_{w,t}(5,15)/d_{w,p}(15)=2$ and the gate ratio $G_t(5)/G_p(15)\approx 24\%$. Thus, with the total number of reuploads, $L$, taken to be three for both models, the aCLS models have a lower gate count than all PDR models.

\bibliographystyle{JHEP}
\bibliography{ref}

\end{document}